\documentclass[a4paper,fleqn,usenatbib]{mnras}

\usepackage{newtxtext,newtxmath}

\usepackage[T1]{fontenc}
\usepackage{ae,aecompl}

\usepackage{graphicx}	
\usepackage{url}
\usepackage{color}

\usepackage{epstopdf}

\newcommand\figwidth{1.01}

\newcommand\Msun{{\rm\thinspace M$_\odot$}}
\newcommand\Mstar{{\rm\thinspace M$_*$}}
\newcommand\Gpc{{\rm\thinspace Gpc}}
\newcommand\erg{{\rm\thinspace erg}}
\newcommand\s{{\rm\thinspace s}}
\newcommand\cm{{\rm\thinspace cm}}
\newcommand\km{{\rm\thinspace km}}
\newcommand\kms{\hbox{$\km\s^{-1}\,$}}
\newcommand{\magg}{\rm\thinspace mag}
\newcommand\nwig{225,415}
\newcommand\nwigw{220,311}
\newcommand\nwigone{110,138}
\newcommand\nmass{184,520}
\newcommand\NUV{\rm NUV}
\newcommand\FUV{\rm FUV}

\include{jdf}

\newcommand\deltabf{}
\newcommand\deltabfz{}
\newcommand\vrc{}
\newcommand\vrd{}

\title[The WiggleZ Dark Energy Survey: Final Data Release]{The WiggleZ Dark Energy Survey: Final Data Release and the Metallicity of UV-Luminous Galaxies}

\author[M. J. Drinkwater et al.]{Michael J.\ Drinkwater,$^1$\thanks{E-mail: m.drinkwater@uq.edu.au}
    Zachary J.\ Byrne,$^1$   
    Chris Blake,$^2$ 
    Karl Glazebrook,$^2$ \newauthor  
    Sarah Brough,$^3$  
    Matthew Colless,$^4$  
    Warrick Couch,$^5$     
    Darren J.\ Croton,$^2$ \newauthor  
    Scott M.\ Croom,$^6$         
    Tamara M.\ Davis,$^1$   
    Karl Forster,$^7$       
    David Gilbank,$^8$     \newauthor  
    Samuel R.\ Hinton,$^1$
    Ben Jelliffe,$^6$       
    Russell J.\ Jurek,$^{1}$  
    I-hui Li,$^9$   \newauthor         
    D.\ Christopher Martin,$^7$  
    Kevin Pimbblet,$^{10,11}$     
    Gregory B.\ Poole,$^{12}$  \newauthor  
    Michael Pracy,$^{6}$   
    Rob Sharp,$^{4}$         
    Jon Smillie,$^{13}$  
    Max Spolaor,$^{14}$
    Emily Wisnioski,$^{15}$  \newauthor  
    David Woods,$^{16}$   
    Ted K.\ Wyder$^{17}$ and 
    H.K.C.\ Yee$^{9}$      
\\
  $^1$ School of Mathematics and Physics, University of Queensland, Brisbane, QLD 4072, Australia \\   
  $^2$ Centre for Astrophysics and Supercomputing, Swinburne University of Technology, P.O. Box 218, Hawthorn, VIC 3122, Australia \\ 
  $^3$ School of Physics, University of New South Wales, NSW 2052, Australia \\   
  $^4$ Research School of Astronomy and Astrophysics, Australian National University, Canberra, ACT 2611, Australia \\ 
  $^5$ Australian Astronomical Observatory, P.O. Box 915, North Ryde, NSW 1670, Australia \\
  $^6$ Sydney Institute for Astronomy, School of Physics, University of Sydney, NSW 2006, Australia \\ 
  $^7$ California Institute of Technology, MC 278-17, 1200 East California Boulevard, Pasadena, CA 91125, USA \\ 
  $^8$ South African Astronomical Observatory, PO Box 9, Observatory, 7935 South Africa \\
  $^9$ Department of Astronomy and Astrophysics, University of Toronto, 50 St.\ George Street, Toronto, ON M5S 3H4, Canada \\
  $^{10}$ E.A.Milne Centre for Astrophysics, University of Hull, Cottingham Road, Kingston-upon-Hull, HU6 7RX, UK \\ 
  $^{11}$ School of Physics and Astronomy, Monash University, Clayton, VIC 3800, Australia  \\ 
  $^{12}$ School of Physics, University of Melbourne, VIC 3010, Australia \\
  $^{13}$ National Computational Infrastructure, The Australian National University 143 Ward Road, Acton, ACT 2601, Australia\\
  $^{14}$ NASA Independent Verification and Validation,  Jon McBride Software Testing \& Research Laboratory, Fairmont, WV 26554, USA\\
  $^{15}$ Max Planck Institut f\"{u}r extraterrestrische Physik, Giessenbachstra$\beta$e, D-85748 Garching, Germany\\
  $^{16}$ Department of Physics and Astronomy, University of British Columbia, 6224 Agricultural Road, Vancouver, BC V6T 1Z1, Canada \\ 
  $^{17}$ Apigee Corporation, 10 Almaden Blvd, San Jose CA 95113, USA
}

\date{Accepted 2017 November 14. Received 2017 October 19; in original form 2016 October 19}

\pubyear{2018}
\volume{474}

\hypersetup{draft}

\begin{document}
\label{firstpage}
\pagerange{\pageref{firstpage}--\pageref{lastpage}}
\pagerange{4151-4168}
\maketitle

\begin{abstract}
{\deltabf The WiggleZ Dark Energy Survey measured the redshifts of over 200,000 UV-selected ($\NUV<22.8$  mag) galaxies on the Anglo-Australian Telescope. The survey detected the baryon acoustic oscillation signal in the large scale distribution of galaxies over the redshift range $0.2<z<1.0$, confirming the acceleration of the expansion of the Universe and measuring the rate of structure growth within it. Here we present the final data release of the survey: a catalogue of  \nwig\  galaxies and individual files of the galaxy spectra. We analyse the emission-line properties of these UV-luminous Lyman-break galaxies by stacking the spectra in bins of luminosity, redshift, and stellar mass. The most luminous ($-25 \magg <M_{\rm FUV}<-22 \magg$) galaxies have very broad H$\beta$ emission from active nuclei, as well as a broad second component to the [OIII] (495.9 nm, 500.7 nm) doublet lines that is blue shifted by 100 \kms, indicating the presence of gas outflows in these galaxies. The composite spectra allow us to detect and measure the temperature-sensitive [OIII] (436.3 nm) line and obtain metallicities using the direct method. The metallicities of intermediate stellar mass ($8.8<\log(M_*/M_\odot)<10$) WiggleZ galaxies are consistent with normal emission-line galaxies at the same masses. In contrast, the metallicities of high stellar mass ($10<\log(M_*/M_\odot)<12$) WiggleZ galaxies are significantly lower than for normal emission-line galaxies at the same masses. This is not an effect of evolution as the metallicities do not vary with redshift; it is most likely a property specific to the extremely UV-luminous WiggleZ galaxies.}
\end{abstract}

\begin{keywords}
surveys -- 
galaxies: abundances -- 
galaxies: photometry -- 
galaxies: starburst -- 
cosmology: observations --
ultraviolet: galaxies
\end{keywords}



\section{Introduction}
\label{sec-intro}

The WiggleZ Dark Energy Survey  is a spectroscopic survey of UV-selected emission-line galaxies over a volume of 1 $\Gpc^3$ \citep[][Paper 1 hereafter]{Drinkwater2010}. The survey was primarily designed as a cosmological experiment to test models of dark energy across redshifts ranging from $z=0.1$ to $z=1$. This corresponds to the epoch where the Universe transitions from deceleration to acceleration in the `standard' $\Lambda$CDM cosmological model. 

The key cosmology results from the project have been published in a series of papers.  We determined a comprehensive measurement of the distance-redshift relation using baryon acoustic oscillations (BAOs) as a standard ruler \citep{Blake2011b}, including the first such measurements in the $z > 0.5$ Universe.  Our final BAO results, including the reconstruction of the acoustic peak based on the estimated displacement field, {\deltabf correspond to distance measurements with relative uncertainties of 4.8, 4.5, and 3.4 per cent} in overlapping redshift slices at $z =$ 0.44, 0.6, and 0.73 \citep{Kazin2014}. We extended that analysis to measure distances along and perpendicular to the line of sight in the same redshift bins \citep{Hinton2016}. By fitting redshift-space distortions in the clustering pattern we measured the growth-rate of structure in the range $0.2 < z < 1$ including its degeneracy with the Alcock-Paczynski distortion \citep{Blake2011a, Blake2012, Contreras2013}.  Our simultaneous constraints on the expansion and growth history are consistent with the predictions of a $\Lambda$CDM cosmological model with a cosmological constant dark energy component and large-scale gravity described by General Relativity, and constitute a new and precise test of that model.

We also presented cosmological-parameter fits to the shape of the WiggleZ galaxy power spectrum in combination with {\deltabf cosmic microwave background} (CMB) measurements \citep{Parkinson2012}, with a particular focus on placing limits on the neutrino mass \citep{Riemer2012}.  The combination of {\it Planck} CMB and WiggleZ data places an upper limit of 0.15 eV (95 per cent confidence) on the sum of the masses of the neutrino species \citep{Riemer2013a, Riemer2013b}.  We have released the cosmological data products in the form of observed samples and matched random samples, the measured BAO correlation functions and covariance matrices, and the measured power spectrum and window functions within a CosmoMC module {\deltabf \citep{Lewis2002}}.  These cosmology data products may all be accessed at \url{http://www.smp.uq.edu.au/wigglez-data} \citep[see][]{Parkinson2012}.  The determination of the survey selection function which underpins these measurements was described by \citet{Blake2010}.

The WiggleZ dataset has been used for several other studies of large scale structure.  These include demonstrating that the fractal properties of the galaxy distribution transition towards homogeneity on large scales in the manner predicted by the $\Lambda$CDM model \citep{Scrimgeour2012}; producing the first constraints on the position of the turnover in the large-scale power spectrum and its application as a standard ruler based on the epoch of matter-radiation equality \citep{Poole2013}; quantifying the 3-point correlation function and using its shape-dependence to measure the linear and non-linear galaxy bias properties and hence recover new measurements of the growth factor $\sigma_8(z)$ at high redshift \citep{Marin2013}; and describing the topological structure through Minkowski functionals, including the first distance measurements based on cosmic topology \citep{Blake2014}.  

In order to reach redshifts up to $z\approx 1$ in reasonable exposure times on the 3.9 m Anglo-Australian Telescope we used strong emission-line galaxies as the WiggleZ survey targets. The targets were selected from UV imaging obtained with the {\it Galaxy Evolution Explorer} satellite \citep[{\it GALEX};][]{Martin2005}. A relatively bright flux limit combined with the very large volume of the survey created a sample of the most UV-luminous galaxies in the local universe \citep{Jurek2013}. We have measured the intrinsic properties of these extreme galaxies \citep{Wisnioski2011,Jurek2013}, as well as their intrinsic alignments with the large-scale density field \citep{Mandelbaum2011} and stellar masses \citep{Banerji2013}.

In Paper 1 we concluded that the majority of WiggleZ galaxies were strongly star forming based on emission-line diagnostics. However, \citet{Jurek2013} found that the luminosity functions of the WiggleZ galaxies had an excess over standard Schechter function fits at high luminosity that might represent a contribution from active galactic nuclei (AGN). We revisit the AGN fraction in this paper using the larger sample and the stacking of spectra to achieve higher signal-to-noise {\deltabf (S/N)} spectra to measure line ratios and line widths. 


The stacked spectra also allow us to detect the temperature-sensitive weak [OIII] (436.3nm) emission line and hence measure the galaxy metallicities using the direct method \citep[e.g.][]{Peimbert1969}. The {\deltabf gas} metallicity of a galaxy depends on both the metal production from stars and the bulk movement of gas, so the relationship between the stellar mass and metallicity provides crucial constraints on galaxy evolution models. The galaxy stellar mass-metallicity relation, a positive correlation between stellar mass and metallicity, was demonstrated for local galaxies ($z\approx 0.1$) by \citet{Tremonti2004} using a Sloan Digital Sky Survey (SDSS) sample of 53,000 star forming galaxies. Several subsequent studies \citep[e.g.][]{Ellison2008,Mannucci2010,LaraLopez2010} have found that galaxies with high star formation rates have a lower metallicity than other galaxies at the same stellar mass. {\deltabf The role of star formation was included by \citet{Mannucci2010,Mannucci2011} in a `fundamental metallicity relation' which predicts metallicity as a function of both stellar mass and star formation rate. \citet{LaraLopez2010,LaraLopez2013} suggest that star-forming galaxies can be described by a thin `fundamental plane' in the three-dimensional space defined by metallicity, star formation rate and stellar mass which does not vary significantly with redshift. However, other studies of SDSS galaxy samples have reported that, although star formation rate can affect metallicity, its inclusion does not significantly reduce the scatter in the relation \citep{Yates2012,Salim2014}.


{\vrc Many studies have sought to determine if the mass-metallicity relation evolves with redshift. Several samples, with redshifts as high as $z=1.6$, reveal a consistent decrease in metallicity with redshift at a given stellar mass \citep{Yabe2012,Yabe2014,Zahid2014,delosReyes2015,Yabe2015,Ly2016a}. This evolution was parameterised by \citet{Ly2016a} with the characteristic metallicity decreasing according to $(1+z)^{-2.3}$.

In contrast, the combined relation between metallicity, stellar mass and star formation rate --- the fundamental plane (FP) or the fundamental metallicity relation (FMR) described above --- is not generally found to evolve. \citet{LaraLopez2010} found no change in their FP relation out to high redshifts ($z =$ 0.85, 2.2, 2.5). Similar conclusions of no evolution in the FMR were reached by \citet{Hunt2012,Ly2014,Ly2015,Yabe2015}, although \citet{Ly2015} noted a high intrinsic dispersion in the relation for the low-mass, high star forming galaxies they measured at $z \approx 0.8$. However, there are some departures from the FP and FMR relations at high redshifts. \citet{Hunt2012} report low-metallicity outliers in the form of luminous compact galaxies ($z\approx 0.3$) and Lyman break galaxies ($z>1$). In particular, \citet{Zahid2014} claim to detect evolution in the FMR in a sample of $z\approx 1.6$ galaxies, with metallicities below the FMR for galaxies of mass below $10^{10}$\Msun.}

This range of results at higher redshifts partly reflects {\vrd the difficulty} in making metallicity measurements at high redshift, as noted by \citet{Andrews2013} but it may also indicate real differences between the evolution (or otherwise) of different galaxy types. \citet{Troncoso2014} report strong evolution to lower metallicity in a sample of star-forming galaxies at $z=3.5$, but still note that they may be measuring a different type of galaxy to the low redshift samples. \citet{Troncoso2014} stress the need for more independent galaxy samples.} The WiggleZ galaxies provide a new independent measurement of metallicity {\deltabf at intermediate ($z \approx 0.6$) redshifts}  by virtue of their Lyman break selection. In this paper we measure the metallicities of stacked WiggleZ galaxy spectra using the approach of \citet{Andrews2013}, extending their work to higher masses, star formation rates and redshifts.


In this paper we now present the full catalogue of all successful observations from the WiggleZ survey, plus data files of all the spectra. The final catalogue contains \nwig\  redshifts, a 125 per cent increase over the \nwigone\ redshifts presented in Paper 1. This is different from the cosmological data presented by \citet{Parkinson2012} in that it contains all objects observed, not just the galaxies used for the cosmological analysis (which were subject to more stringent selection criteria). We also provide more extensive observational data for the objects. We start this paper with an overview of the survey design, including descriptions of the target selection and the observations in Section~\ref{sec-design}. In Section~\ref{sec-results}, we characterise the WiggleZ galaxies, notably the properties we can measure from stacked spectra. We discuss the implications of these results in Section~\ref{sec-discuss}. Following this, in Section~\ref{sec-data} we describe the data products in our final public data release accompanying this paper. We summarise all the results in Section~\ref{sec-summary}. A standard cosmology of $\Omega_m = 0.3$, $\Omega_{\Lambda} = 0.7$, $h= 0.7$ is adopted throughout this paper.

\section{WiggleZ Dark Energy Survey Design and Analysis Methods}
\label{sec-design}

In this section we review the design of the WiggleZ Survey and describe the analysis methods used in this paper. Full details of the design and calibration are given in Paper 1.

\subsection{Survey fields}

\begin{figure*}
\includegraphics[width=0.95\linewidth]{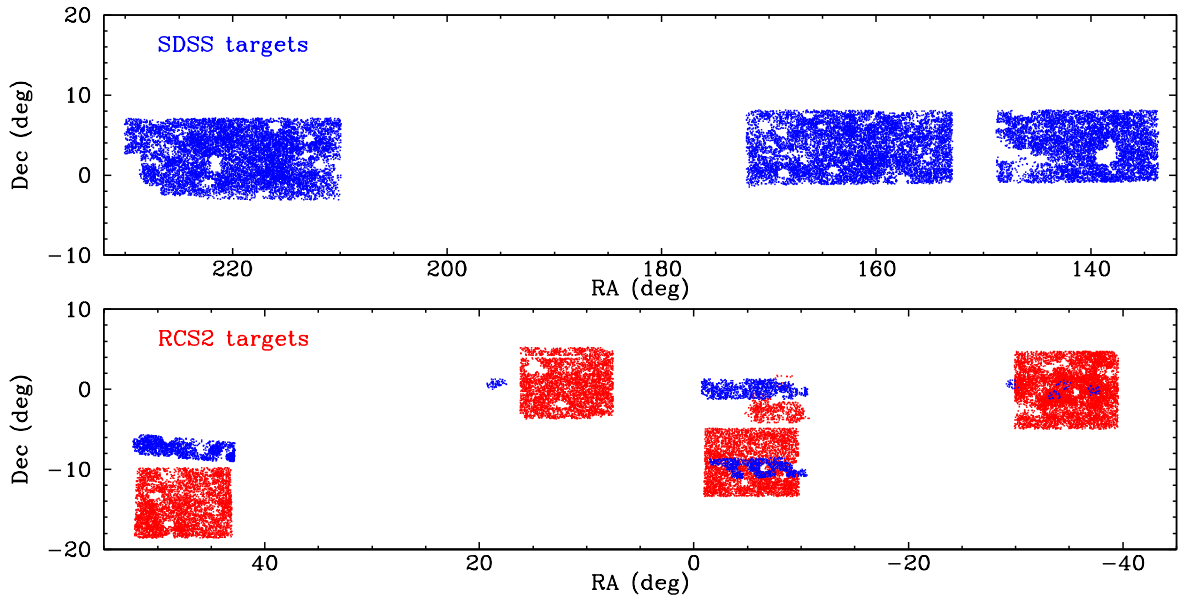}
\caption{The distribution on the sky of observations in the seven WiggleZ survey
    regions. The colours of the points indicate if the targets used the SDSS (blue) or RCS2 (red) optical data. The circular gaps in the distribution are mostly due to bright stars which had to be avoided by the {\it GALEX} observations.}
\label{figsurvey}
\end{figure*}

The WiggleZ survey was conducted in seven equatorial fields covering a total area of approximately 1000 deg$^2$, as illustrated in Fig.~\ref{figsurvey} and listed in Table~\ref{tab-surveyhigh}. The arrangement of the fields was chosen to allow year-round observing, whilst keeping each field large enough that its smallest dimension was three times as large as the 100 $h^{-1}$ Mpc baryon acoustic oscillation scale. 

\begin{table}
\caption{Survey Regions: Final Areas}
\label{tab-surveyhigh}
\begin{tabular}{rrrrrrr}
\hline
Name & RA$_{\rm min}$ & RA$_{\rm max}$ & Dec$_{\rm min}$ & Dec$_{\rm max}$ & Area& {\deltabf $N_{t}$}\\
     & (deg)     & (deg)     & (deg)     & (deg)   & (deg$^2$)  \\
\hline
  0-hr  &  350.1 & 359.1  &  $-13.4$ & $+1.8$ &  135.7 &  {\deltabf 48409} \\
  1-hr  &      7.5 &   16.5  &    $-3.7$ & $+5.3$ &   81.0  &  {\deltabf 34507} \\
  3-hr  &    43.0 &   52.2  &  $-18.6$ &  $-5.7$ &  115.8 &  {\deltabf 51534} \\  
  9-hr  &  133.7 & 148.8  &    $-1.0$ & $+8.0$ &  137.0 &  {\deltabf 55575} \\
 11-hr &  153.0 & 172.0  &    $-1.0$ & $+8.0$ &  170.5 &  {\deltabf 70644} \\
 15-hr &  210.0 & 230.0  &    $-3.0$ & $+7.0$ &  199.6 &  {\deltabf 80875} \\
 22-hr &  320.4 & 330.2  &    $-5.0$ & $+4.8$ &   95.9  &  {\deltabf 62491} \\
\hline
\end{tabular}
{\deltabf Note: $N_{t}$ is the number of potential targets for spectroscopic observations in each region.}
\end{table}

\subsection{Photometry}
\label{sec-galex}

The WiggleZ galaxies were primarily selected on the basis of ultra-violet photometry {\deltabf from the wide-field {\it GALEX} Medium Imaging Survey, which was extended to cover the WiggleZ fields.}  We used data from both {\it GALEX} bands: the FUV (135--175\,nm), and the NUV (175--275\,nm). {\deltabf The NUV magnitude corresponds to the flux through an elliptical aperture scaled to twice the Kron radius of each source and the FUV magnitude is from a fixed 12-arcsec circular aperture at the location of the NUV detection. We correct both the NUV and FUV magnitudes for Galactic dust extinction \citep[see Paper 1, and][for details]{Morrissey2007}.}

The {\it GALEX} exposure times (around 1500 seconds) were relatively short, so the {\deltabf number density} of WiggleZ targets at our survey limit of NUV$<22.8$ mag  is significantly incomplete (see Paper 1). The FUV data in particular proved to contain limited numbers of significant detections \citep[see Paper 1,][]{Jurek2013}.
{\deltabf Our survey limit of NUV$<22.8$ approximately corresponds to a 3-sigma detection limit, but this varies with the amount of dust correction. We therefore applied a detection limit to the raw (uncorrected for dust) NUV fluxes of $S/N_{}>3$ (see Paper 1).}

We combined the {\it GALEX} UV data with optical photometry to improve our target selection, as well as to provide more accurate positions for the spectroscopic observations. The optical data were taken from the {\deltabf fourth data release\footnote{\deltabf This was the current version of the SDSS when we designed the WiggleZ survey.}} of the Sloan Digital Sky Survey \citep[SDSS;][for RA = 130--230 deg]{Adelman2006} and from the Canada-France-Hawaii Telescope Second Red-sequence Cluster Survey \citep[RCS2;][{\vrc for RA = $-$40 to 55 deg}]{Gilbank2011}. Note that we used a preliminary version of the RCS2 photometry. {\deltabf We used a 2.5 arc second radius to match the {\it GALEX} and optical positions, giving a 95 per cent confidence in the SDSS matches and a 90 per cent confidence in the RCS2 matches (see Paper 1).} Fig.~\ref{figsurvey} shows all the objects in the final catalogue colour-coded by the origin of the optical photometry. Note that the {\vrd South Galactic Pole} ({\vrd RA = $-$40 to 55 deg}) fields use RCS2 photometry except for a small amount of SDSS data for regions outside the RCS2 fields or for early test observations. 

We checked the astrometry of all our optical data against the 2 Micron All-Sky Survey \citep[2MASS,][]{Skrutskie2006} catalogue. The SDSS data were all consistent with the 2MASS astrometry, but we found small offsets between the RCS2 reference frame \citep[defined by the United States Naval Observatory B (USNO-B) catalogue,][]{Monet2003} and 2MASS. {\deltabf We have corrected the RCS2 galaxy positions (unlike in Paper 1) so that all positions in our final catalogue are consistent with the 2MASS {\vrc astrometry}. The small average offsets we corrected for in each RCS2 field are listed in Table~\ref{tab-offset}.}

\begin{table}
\caption{Mean Astrometry Offsets in RCS2 Regions.}
\label{tab-offset}
\begin{tabular}{rrr}
\hline
Name & RA$_{\rm USNO}-$RA$_{\rm 2MASS}$ & Dec$_{\rm USNO}-$Dec$_{\rm 2MASS}$ \\
     & (arc seconds)     & (arc seconds)  \\
\hline
  0-hr &  $-0.112\pm 0.003$ &  $0.117\pm 0.003$  \\
  1-hr &  $-0.133\pm 0.004$ &  $0.102\pm 0.003$  \\
  3-hr &  $-0.141\pm 0.002$ & $-0.026\pm 0.002$  \\
 22-hr &  $-0.031\pm 0.003$ &  $0.137\pm 0.003$  \\
\hline
\end{tabular}

Note: The WiggleZ coordinates in the RCS2 regions are based on USNO astrometry. The uncertainty given for each mean offset is the standard error of the mean.
\end{table}

\subsection{Galaxy Selection}
\label{sec-selection}

The WiggleZ sample was principally defined by a (Galactic extinction-corrected) limiting magnitude of 22.8 in the {\it GALEX} NUV band. {\deltabf We then used two colour terms to limit the sample to $z>0.5$ emission-line galaxies. We required $\FUV-\NUV>1$ (the Lyman break passes between these filters at $z\approx 0.5$) and $-0.5<\NUV-r<2$ to select emission-line galaxies. This selection is illustrated as a colour-colour diagram {\vrc in fig.\ 5 of Paper 1.}} 

{\deltabf 
{\vrc A high median redshift was essential for our cosmology experiments, so } we applied additional criteria to the optical photometry to further increase the fraction of high-redshift ($z>0.5$) galaxies. 
\begin{enumerate}
\item We rejected targets likely to be low-redshift ($z<0.5$) galaxies according to their colours, as described in the second part of Table~\ref{tab-selection}. Note that for the SDSS regions we restricted this rejection to the brighter targets (imposing $g$ and $i$ limits) with reliable colours. 
\item We imposed a bright limit on the optical $r$-band ($r<20$) to avoid low-redshift {\deltabf ($z<0.5$)}  galaxies. When combined with the allowed $\NUV-r$ colours, this resulted in a bright UV magnitude limit of $\NUV>19.5$, although this was not a formal selection criterion. 
\end{enumerate}

\noindent  This additional selection removed about half the remaining low-redshift ($z<0.5$) galaxies from the sample as shown {\vrd in fig.\ 7 of Paper 1.} We list all the selection criteria in Table~\ref{tab-selection} (see Paper 1 for details). }

{\deltabf We did not select by image morphology (e.g., avoiding objects classified as stars) because very few (only 0.7 per cent of the sample; see Sec.~\ref{sec-spec}) stars satisfied the combination of all the photometric selection criteria.}

\begin{table}
\caption{Photometric selection criteria for WiggleZ galaxies.}
\label{tab-selection}
\begin{tabular}{ll}
\hline
Criterion  &  Values     \\
\hline
\multicolumn{2}{l}{\em (i) Select targets satisfying:}\\ 
Magnitude &  $\NUV<22.8$  mag  \\
Magnitude &   $20 < r < 22.5$  mag\\
Colour    &  $(\FUV-\NUV) > 1$  mag or no FUV  \\
Colour    & $-0.5 < (\NUV-r) < 2$   mag\\
Signal    &  $S/N_{\NUV}>3$ \\
Optical Position  & matches within 2.5 arc seconds \\
\hline
\multicolumn{2}{l}{\em (ii) Reject targets satisfying {\deltabf all the following:}}\\ 
SDSS regions: & $g < 22.5$ {\deltabf and} $i < 21.5$  {\deltabf and}\\
  & $(r-i) < (g-r-0.1) $ {\deltabf and} $ (r-i) < 0.4$  mag\\
RCS2 regions: & $(g-r) > 0.6$ {\deltabf and  }
$(r-z) < 0.7(g-r)$  mag\\
\hline
\end{tabular}
\end{table}

We also prioritised the order in which galaxies were observed according to the rules given in Paper 1. The main effect was to observe fainter objects (according to their optical $r$ band magnitude) first as they were more likely to have higher redshifts. This means that our spectroscopic completeness is a function of $r$-band flux.

If we were unable to allocate all the fibres to high-priority WiggleZ targets, we observed targets from a small number of `spare fibre' projects. In total these amounted to less than 2 per cent of all the final catalogue measurements (see Table~\ref{tab-targetcodes}). The three projects were:
\begin{enumerate}
\item Objects exhibiting short-term UV variability, as measured by the {\it GALEX} satellite (code `G').
\item Candidate galaxy cluster members drawn from the RCS2 survey (`X'), \citep[see][]{Li2012}.
\item Candidate radio galaxies from the FIRST \citep{Becker1995} survey (`Y'), \citep[see][]{Pracy2016,Ching2016}.
\end{enumerate}

\subsection{Spectroscopy}
\label{sec-spec}

We used the AAOmega floor-mounted spectrograph fed by 392 optical fibres from the 2-degree Field top end of the Anglo-Australian Telescope. We initially used the standard low-resolution observing setup with light feeding the two arms of the spectrograph separated by a dichroic beam splitter at a wavelength of 570 nm. We subsequently obtained a second dichroic centred on 670 nm to increase our ability to identify high-redshift objects. Observations made after 2007 August 1 used the new dichroic. The parameters of both dichroics are given in Table~\ref{tab-aaomega}. We used the 580V and 385R gratings in the blue and red arms of the spectrograph, respectively. Both gave a resolution of $R\approx 1300$ (0.36 nm in the blue and 0.55 nm in the red). We note that the blue spectrograph camera was unavailable for one run (2007 June 8--21) so data from these dates only cover the red half of the spectrum.

\begin{table}
\caption{Spectrograph configurations during the survey.}
\label{tab-aaomega}
\begin{tabular}{lrrl}
\hline
Dichroic  &   $\Delta\lambda_{\rm Blue}$  &  
$\Delta\lambda_{\rm Red}$ & Dates \\
\hline
570 nm  &    370--580 &  560--850 & 2006--2007 Apr 17 \\
570 nm  &    N/A      &  560--850 & 2007 Jun 8--21 \\
570 nm  &    370--580 &  560--850 & 2007 Jul 6--9 \\
670 nm  &    470--680 &  650--950 & 2007 Aug 7--2011 Jan 13 \\
\hline
\end{tabular}

Note: for each dichroic the table lists the
observable wavelength range for the blue and red arms of the
spectrograph. All wavelength units are nm. The blue camera was not
available for the 2007 June observations.
\end{table}

The AAOmega spectra were processed using the standard AAO {\sc 2dfdr} pipeline software. This applied throughput corrections to the fibre spectra based on the intensity of the strongest atmospheric emission lines to normalise the spectra before subtracting a sky signal measured from dedicated fibres positioned on regions of blank sky. The data from the two arms of the spectrograph were spliced together, resulting in a single FITS file containing typically 360 object spectra and their corresponding variance (noise) spectra, all about 4500 pixels long. We did not apply a correction for Telluric (atmospheric) absorption. For the public data release we have extracted an individual FITS file (spectrum plus variance) from the original, reduced, multi-spectrum files for each galaxy in the final catalogue.

It is challenging to calculate absolute spectrophotometric calibration of objects measured with fibre systems like AAOmega due to the large aperture corrections. However, it is necessary to remove the instrumental response in order to splice the blue and red spectra. This requires transformation of the spectra from raw counts to calibrated flux units. The transfer function for the AAOmega spectrograph was based on observations of a standard star (EG 21). The calibration was calculated when AAOmega was first commissioned, and then again when the new dichroic was installed. The two transfer functions (for the respective dichroics) were not recalibrated during the WiggleZ survey, but tests show that any long-term variations are small compared to individual errors on single observations \citep{Sharp2013}. {\deltabf For this reason we have not attempted to provide any improved absolute calibration of the spectra.}

The WiggleZ spectra presented here are therefore nominally flux calibrated, measured in units of ${\vrd 10^{-16}} \erg \s^{-1} \cm^{-2}$ \AA$^{-1}$ (and linearly binned in wavelength). However, the overall uncertainty in the calibration of a given spectrum is large (up to a factor of 10) due to the uncertain aperture correction inherent in fibre spectroscopy (noted above). {\vrd We estimated the extent of this scatter by comparing the observed continuum levels in the spectra with the levels predicted from their apparent $m_r$ magnitudes. We found that only 0.1 per cent of spectra were more than 10 times brighter than predicted and only 0.4 per cent were more than 10 times fainter than predicted.}

We measured redshifts for all the galaxies observed using the {\sc runz} software \citep{Saunders2004} originally developed for the 2dF Galaxy Redshift Survey \citep{Colless2001}. This was optimised to measure emission-line redshifts for the WiggleZ data, {\deltabf but also used a range of star and galaxy templates to measure absorption line redshifts.} {\sc runz}  calculated an average correction for telluric (atmospheric) absorption and applied that to each spectrum before the analysis. The {\sc runz} software assigned an automatic redshift to each galaxy, but manual intervention was required to confirm or correct the redshift. Every redshift was checked by one of the authors and assigned a quality code $Q$, as described in Table~\ref{tab-quality}.

\begin{table*}
\caption{Redshift Quality Codes.}
\label{tab-quality}
\begin{tabular}{lrrl}
\hline
Q & {\deltabf N} & {\vrc Fraction} & Definition \\
\hline
1 & 0          &     0.0000  & No redshift was possible; highly noisy spectra.\\
2 & 122      &     0.0005  & An uncertain redshift was assigned.\\
3 & 76725  &     0.3404  & A reasonably confident redshift; if based on [OII]  alone, the doublet is resolved or partially resolved.\\
4 & 120371 &    0.5340  & A redshift that has multiple (obvious) emission  lines all in agreement.\\
5 &  28192  &    0.1251  & An excellent redshift with high S/N that may be  suitable as a template.\\
\hline
\end{tabular}

{\deltabf Note: columns 2 and 3 give the number and fractions of all the {\em final} }catalogue objects in each category. Most objects with a quality lower than 3 are excluded from the catalogue: those remaining were originally classified as stars and remain for compatibility with earlier versions of the survey.
\end{table*}

As we describe in Paper 1, we originally used an additional redshift quality code with a value of $Q=6$. This was intended to flag a small number of white dwarf stars observed for calibration purposes (see Paper 1), but it was also applied in some cases to extragalactic objects when they had clear features of active galactic nuclei in their spectra. For the current catalogue we have reclassified all the objects originally marked $Q=6$, purely according to the reliability of their redshift\footnote{{\deltabf {\sc runz} can assign redshifts to stars as it has a range of stellar templates, including white dwarfs.}}, assigning them codes in the range $Q=$2-5.

We have excluded objects which were originally assigned a quality code $Q<3$ from this final data release. However, we have included the small number (122) of reclassified objects which now have a quality code of $Q=2$. This is to provide compatibility with earlier versions of the catalogue (e.g. Paper 1).

The only way to identify Galactic objects in the final catalogue is therefore by redshift. We estimated the standard deviation of the velocities of the Galactic stars in the catalogue (after clipping to $|z|<$0.004 = 1200 km s$^{-1}$) as $\sigma_z=0.00081$ ($\sigma_v=248 \km\s^{-1}$). Adopting a 3-sigma cutoff of $z<0.0024$ we identify 1574 Galactic objects (0.7 per cent of the total) in the final catalogue according to redshift. 


\subsection{Redshift Reliability and Distribution}
\label{sec-reliable}

{\deltabf We repeated observations of approximately 10 000 randomly-selected galaxies during the survey to test both the precision and reliability of our redshift measurements. We considered redshifts to be reliable (the correct lines identified) if the repeated redshifts agreed to within $|\Delta z|<0.002$ (375 \kms at our median redshift of $z=0.6$). For the reliable redshift pairs, we calculated the uncertainty as the standard deviation of $\Delta z$ scaled by $1/\sqrt{2}$. We present the precisions and the fractions of reliable redshifts for the different quality categories in Table~\ref{tab-redshift}.} The table shows very similar results to our original analysis in Paper 1, notably that the reliability increases substantially from single-line identifications ($Q=3$) to multiple-line identifications ($Q>3$). The redshift uncertainty is 48 \kms or better in all categories. 
{\deltabf We calculated the redshift reliability for different values of $r$-band magnitude, but found there was no significant difference. This is presumably because the redshifts are determined from emission lines and the r-band magnitude is not a strong predictor of the emission line strength for these UV-selected galaxies.}


\begin{table}
\caption{Reliability and Uncertainty of Redshift Measurements.}
\label{tab-redshift}
\begin{tabular}{cccrrrr}
\hline
Quality  &  Reliability & \multicolumn{2}{c}{Uncertainty}  \\
    $Q$  &      (per cent)        & $\sigma_z$ & $\sigma_v$ (\kms)   \\
\hline
3  &  84.2   &  0.000 27  & 47.9 \\
4  &  98.2   &  0.000 20  & 37.8 \\
5  &  99.7   &  0.000 17  & 32.9 \\
\hline
\end{tabular}

{\deltabf Notes: reliability is the fraction of repeated measurements giving similar ($|\Delta z|<0.002$) redshifts. The 
uncertainties $\sigma_z, \sigma_v$ were obtained by scaling the internal standard
deviations of $\Delta z$  by $1/\sqrt{2}$. The velocity differences for each measurement were
calculated as $\Delta v = c \Delta z / (1 + z)$. See Sec.~\ref{sec-reliable}.}
\end{table}

\begin{figure}
\includegraphics[width=\figwidth\linewidth]{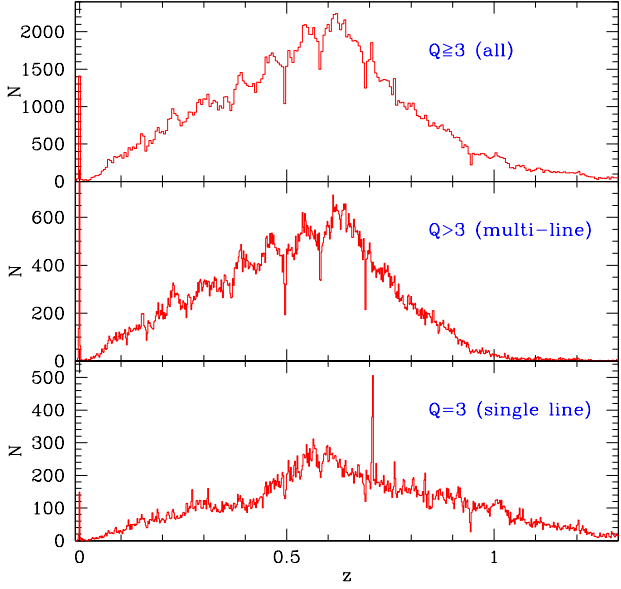}
\caption{The distribution of galaxy redshifts in the final WiggleZ sample. Top panel: the full sample. Middle panel: galaxies identified by multiple lines  (quality flag $Q>3$). Bottom panel: galaxies identified by a single line or with low signal-to-noise ($Q=3$). The distributions show sharp peaks due to sky lines interfering with the redshift measurements; the peaks near redshift $z=0$ are due to Galactic stars.
}
\label{fig-quality}
\end{figure}

We show the final redshift distribution of the WiggleZ galaxies in Fig.~\ref{fig-quality}, for both the full sample (top) and single and multi-line redshifts separately. The increased sample size reveals some features not evident in the early data presented in Paper 1.

In the single line (or noisy) identifications (Q=3) there is a distinct peak at $z=0.708$. This redshift corresponds to the [OII] emission line falling at the position of a weak sky line at 636.2nm (as noted in Paper 1). The range of the peak is $z=0.7064$-0.7084. In this redshift range there are 678 $Q=3$ redshifts, but only 185 are expected. Allowing for this systematic error, we find that $70\pm10$ per cent of the $Q=3$, $0.7064<z<0.7084$ redshifts are likely to be spurious. {\deltabf If we remove objects in this redshift range, the reliability for $Q=3$ measurements increases by 1 per cent.}

The multi-line ($Q>3$) redshifts display three narrow dips at redshifts $z=0.496$, 0.582, and 0.690. These correspond to the [OII] emission line falling on strong sky lines at wavelengths 557.7 nm,  589.3 nm and 630.0 nm respectively. The appearance of these dips in the $Q>3$ redshift sample indicates that these strong sky lines affect the use of [OII] as a secondary line to confirm redshifts. {\deltabf These sources can still be identified at these redshifts since the H$\beta$ and/or [OIII] lines can still be detected in the spectrum.}

\subsection{Stellar mass estimates}
\label{sec-masses}
We calculated stellar masses for the WiggleZ galaxies from their UV and optical photometry using the KG04 spectral fitting code \citep{Glazebrook2004} with PEGASE.2 stellar models \citep{Fioc1997,Fioc1999} and a \citet{Baldry2003} initial mass function. This is the same method as \citet{Banerji2013} previously applied to a subset of 40,000 WiggleZ galaxies: they found that the median $1-\sigma$ uncertainty in log(stellar mass) from the fits to the UV plus optical photometry is 0.48 dex. \citet{Banerji2013} found that the estimates improved when infra-red photometry data were included, but this is not available for the full WiggleZ sample we present here. We restricted the fitting to the redshift range $0.3<z<1.3$ \citep[as did][]{Banerji2013} and find mass solutions for 82 per cent (\nmass) of the WiggleZ galaxies. We include these masses in our final catalogue.

\subsection{Stacked spectra and spectral line fitting}
\label{sec-stackfits}

Most WiggleZ spectra have no detectable continuum signal due to the faint magnitudes of these objects. However, we can detect the continuum in stacked spectra thanks to the very large size of the WiggleZ sample. In this section we describe how we calculated and analysed the stacked spectra.

{\deltabf We only averaged spectra that were directly selected for the main WiggleZ survey, excluding the `spare fibre' objects (see Sec.~\ref{sec-selection}). We also excluded any spectra that were flagged as `fringed' in our visual inspection (optical interference patterns; see Paper 1). This left a total of 217,243 spectra available for stacking from the full sample of \nwig\  galaxies. Before stacking the spectra we also masked out small regions around the four strongest night sky emission lines ([OI] 557.7, Na 589.3, [OI] 630.0, and OH 895.8 nm) as there were often large residual errors in the processed spectra at these wavelengths. The spectra were shifted to rest wavelength before stacking. We also rejected any points more than 20 sigma from the mean at each wavelength to remove any remaining bad data, such as poorly removed cosmic ray events. We did not correct the spectra for Galactic reddening before averaging because the correction was negligible at the low extinctions in the WiggleZ fields and the red observed wavelengths.}

\begin{figure*}
\includegraphics[width=\figwidth\linewidth]{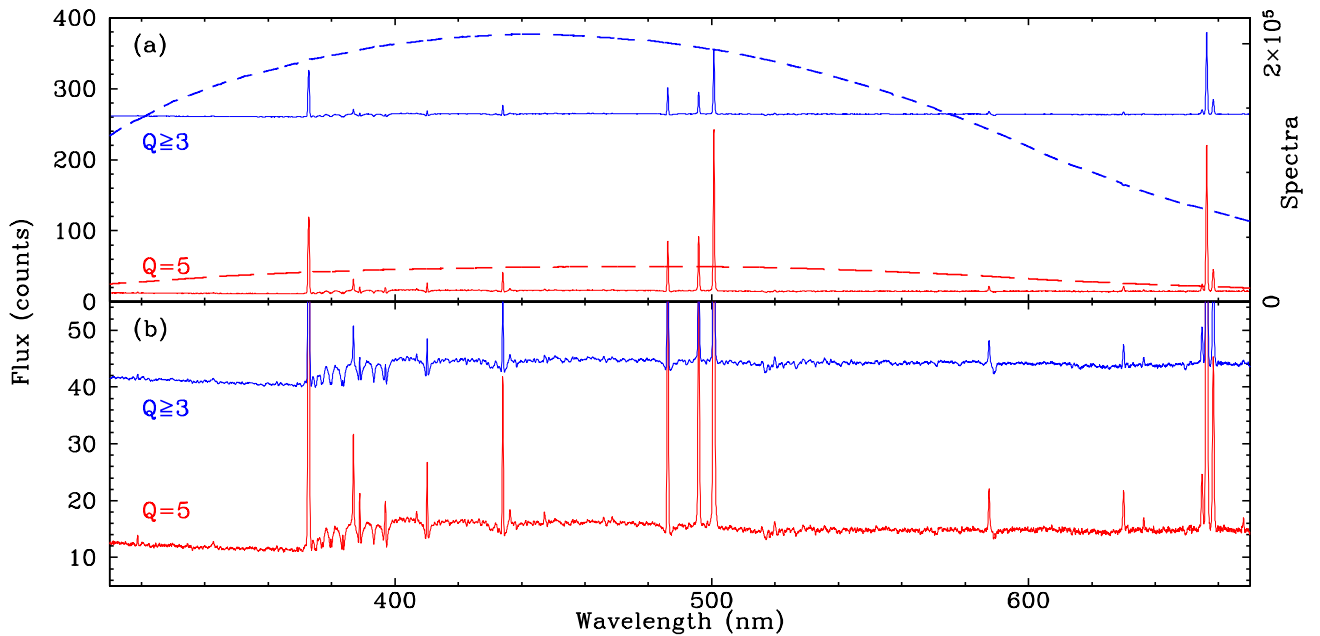}
\caption{Average rest-wavelength spectra of WiggleZ galaxies. The upper panel (a) shows the average of all spectra ($Q\ge 3$, blue) as well as just the high-quality spectra ($Q=5$, red). The rest-wavelength range observed in each spectrum varies with redshift. This is indicated by the dashed lines which show the number of spectra contributing to each average spectrum as a function of wavelength. The lower panel (b) shows the same spectra with the vertical scale enlarged to show the weaker continuum features. The $Q\ge 3$ spectra in both panels are offset by an arbitrary amount.
}
\label{fig-mean1}
\end{figure*}

We present average spectra of the entire sample, and just the high-quality objects (with the redshift  quality flag $Q=5$) in Fig.~\ref{fig-mean1}. Note that the rest-wavelength range observed varies with redshift, so the number of individual spectra contributing to these mean spectra varies with wavelength, as shown by the dashed lines in Fig.~\ref{fig-mean1}. These average spectra display a wealth of weaker features not seen in individual spectra, including both weak emission lines and many continuum features.

\begin{figure}
\includegraphics[width=\figwidth\linewidth]{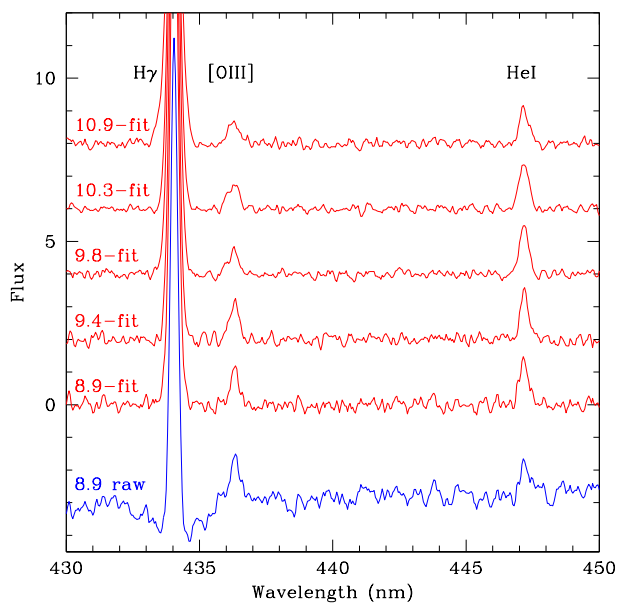}
\caption{Average rest wavelength spectra in the region of the [OIII] 436.3 nm line, calculated for bins of decreasing stellar mass, labelled by their values of log(\Mstar/\Msun). The spectra have had a stellar continuum fit subtracted to enable the detection and measurement of the weak [OIII] line. The lowest mass spectrum is repeated at the bottom without the continuum subtraction to demonstrate how much continuum structure is removed. The structure between the strong lines is not all noise, but contains real spectral features from weak metal lines that are removed by the continuum fitting.}
\label{fig-o3line}
\end{figure}

{\deltabf We fitted and subtracted the stellar continuum from each average spectrum before measuring the emission lines. This is particularly important to remove the stellar Balmer absorption near the weak [OIII] (436.3 nm) line. We used the same approach as
 \citet{Andrews2013}. We used the {\sc Starlight} \citep{Cid2005} program to fit the stellar continuum, adopting the \citet{Cardelli1989} extinction law and using the MILES \citep{Sanchez2006,Falcon2011} spectral templates. We fitted the three regions of interest ([OII] 372.7 nm, $H\gamma$ with [OIII] 436.3 nm, and H$\beta$ with [OIII] 495.9, 500.7 nm) separately to obtain the best models of their stellar continua. Each region was then renormalised to its original continuum flux to give correct line ratios between regions. We show several stacked spectra after subtracting the continuum fits in the  [OIII] 436.3 nm region in Fig.~\ref{fig-o3line}. We also show the raw average spectrum in one case to demonstrate the amount of continuum structure removed by the fitting process.}

We fitted Gaussian profiles to the emission lines (using Chi-square minimisation) to measure their fluxes and widths. We applied both one and two-component fits to each strong emission line. The two component fits for [OIII] and H$\beta$ help identify if winds or broad AGN components are present, respectively. The [OII] doublet, when resolved, can constrain the electron density (see Sec.~\ref{sec-width}). We used the Bayes information criterion \citep[BIC;][]{Kass1995} to determine if the two-component fit was preferred. We calculated this criterion as
\begin{eqnarray}
BIC = \chi^2 + k \ln(n),
\end{eqnarray}
where $\chi^2$ is the normal deviation of the model from the data, $k$ is the number of model parameters, and $n$ is the number of data points. In accordance with \citet{Kass1995} we consider the two-component fit is `not' or is `very strongly' preferred if the BIC difference (one-component $-$ two-component) has values of $<6$ or $>10$ respectively (none of the fits had values between 6 and 10). These are indicated as `n' and `Y' in Table~\ref{tab-2fits}. {\deltabf In the case of the broad H$\beta$ line component, it is possible that the stellar absorption might be over-subtracted {\vrc or under-subtracted}, leading to a false broad component. For this reason we only claim detection of a broad H$\beta$  component when the broad line flux is more than 20 per cent of the narrow line flux -- this only applies to the two most luminous samples in Table~\ref{tab-2fits}.}

\begin{figure}
\includegraphics[width=\figwidth\linewidth]{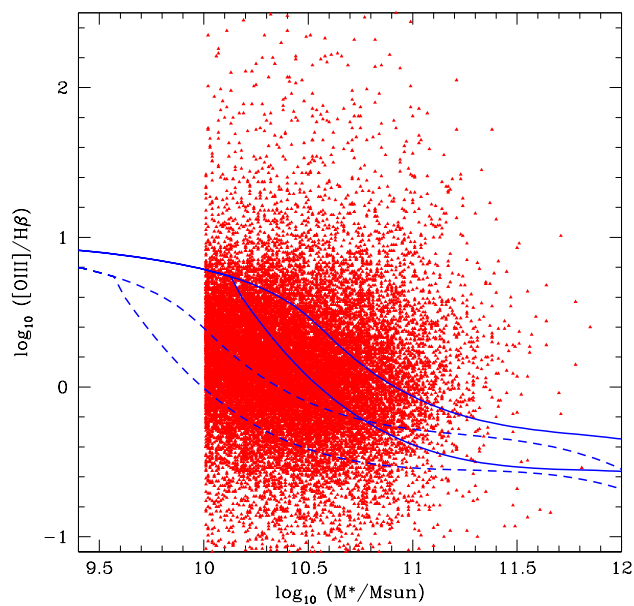}
\caption{{\vrd The mass-excitation (MEx)} diagnostic plot for a sample of WiggleZ galaxies. The points show the [OIII]/H$\beta$ line ratio and stellar mass of galaxies in the bin with  $10<$ log(\Mstar/\Msun) $<12$ and $0.3<z<0.53$ (see Table~\ref{table:lineratios}). The dashed lines show the MEx demarcation lines for low-redshift galaxies \citep{Juneau2014} and the solid lines show these shifted for the line luminosity detection limit of this sample. Galaxies above the upper line are expected to be more than 70 per cent AGN. We classified any galaxies above the upper solid demarcation line and with H$\beta$ line luminosities above the detection limit as probable AGN and excluded them from the metallicity calculations. 7.5 per cent of the galaxies in this bin were removed.}
\label{fig-mex}
\end{figure}

{\vrc
\subsection{Active galactic nuclei in stacked spectra}
\label{sec-mex}

It is important to avoid AGN when stacking spectra because they can bias the metallicity measurements \citep{Andrews2013}. AGN are normally identified using emission line diagnostics  based on the two ratios [OIII]/H$\beta$ and [NII]/H$\alpha$ \citep[BPT][]{Baldwin1981}. This is not possible for most of the WiggleZ sample as the H$\alpha$ [NII]  lines are not observable once the redshift is greater than $z=0.48$. We therefore used the mass-excitation (MEx) diagnostic \citep{Juneau2014} which relies on the [OIII]/H$\beta$ line ratio and stellar mass, both of which are available for all the WiggleZ galaxies (except those with redshift $z<0.3$). 

The MEx diagnostic presented by \citep{Juneau2014} uses the BPT diagnostics of a low-redshift galaxy sample to determine the fraction of galaxies which are AGN as a function of their position in the two-dimensional space defined by the [OIII]/H$\beta$ line ratio and the stellar mass. They define two demarcation lines in the diagram where the AGN fractions are about 0.4 and 0.7. \citet{Juneau2014} show how the demarcation lines shift for other samples as a function of redshift and the emission line detection limit.

We applied the MEx diagnostic to each sample (bins of mass and redshift) to remove probable AGN before stacking the WiggleZ spectra {\vrd for metallicity analysis} as follows.
\begin{enumerate}
\item We fitted the [OIII] doublets and H$\beta$ lines in individual WiggleZ spectra (using the same method as in Sec.~\ref{sec-stackfits} above) where possible. The lines were fitted successfully in 90 per cent of the spectra.
\item We estimated the detection limit of the H$\beta$ line as the median of the detected (log) luminosities in each sample. This was approximately equivalent to the 3-sigma detection level used by \citet{Juneau2014}. We calculated the luminosities of the H$\beta$ lines from their equivalent widths as described in Sec.~\ref{sec-directmetal} below.
\item We shifted the MEx demarcation lines horizontally (a mass offset) to allow for the line luminosity thresholds of our samples, according to the empirical relation determined by  \citet{Juneau2014} (equation B1). We also allowed for luminosity evolution of the line luminosity according to $L*_{H\beta} \propto (1+z)^{2.27}$. The mass offsets for our samples were typically about 0.5 dex, of which the largest line evolution contribution was $-0.04$ dex.
\item We flagged any objects above the (adjusted) upper demarcation line as probable AGN and removed them from the samples. We did not test galaxies with H$\beta$ luminosities below the line detection thresholds, but such objects are unlikely to be AGN. 
\end{enumerate}

{\vrd We did not remove the objects with high FUV luminosities ($M_{\FUV}<-22$ mag) described as likely AGN in Sec.~\ref{sec-fAGN} because these were mostly at higher redshifts than the metalliticy samples or were removed by the MEx criterion; in the two high-redshift bins (see Table~\ref{table:lineratios}), only 2.2 per cent of the selected galaxies had $M_{\FUV}<-22$ mag.} We show an example of the MEx selection applied to one of our samples (from Table~\ref{table:lineratios}) in Fig.~\ref{fig-mex}.
}

\subsection{Direct method metallicity and star formation rate estimation}
\label{sec-directmetal}

There are two common ways of calculating the gas metallicity from galaxy emission-line spectra: the strong emission-line method \citep[e.g.,][]{Pagel1979} and the direct method \citep{Peimbert1969} {\deltabfz which involves direct measurements of electron temperature from weak temperature-dependent oxygen emission lines. We use the direct method in this paper because we can detect the weak [OIII] (436.3 nm) line in our stacked WiggleZ spectra.}
 
{\deltabfz After measuring the emission line fluxes (as described above), we corrected the fluxes for internal reddening by measuring the $H\gamma / H\beta$ Balmer decrement and applying the \citet{Cardelli1989} dust extinction law}. Once the line fluxes for all [OIII] (436.3,495.9 and 500.7 nm) lines were obtained, we estimated the electron temperature from the {\deltabfz auroral to nebular [OIII] flux ratio \citep[as described in equation 2 of][]{Nicholls2014}}. The [OII] temperature diagnostic lines were not observable, so we inferred the [OII] temperature as $T_{\rm OII} = 0.7 \times T_{\rm OIII}+3000 K$ \citep{Campbell1986}. As discussed by \citet{Andrews2013} this relation is reliable for galaxies with high star formation rates such as the WiggleZ objects.

We then used the relative fluxes of [OII] (372.7 nm) and [OIII] (495.9 + 500.7 nm) to the H$\beta$ line, with the electron temperatures to calculate the {\deltabfz $O^+$ and $O^{++}$ abundances, using the relations presented by \citet{Izotov2006} {\vrc (equations 3 and 5).}} We took the total oxygen abundance to be the sum of the  {\deltabf $O^+$ and $O^{++}$}  abundances. We estimated uncertainties in our final metallicity results by using a Monte Carlo simulation of the process, based on our measured uncertainties of the line fluxes. The metallicity measurements are discussed below in Sec.~\ref{sec-massmetal}.

We calculated dust-corrected star formation rates by using the metallicity-dependent H$\alpha$-SFR relation {\vrc described by \citet{Ly2016b} (equation 17).} Given that most WiggleZ spectra do not cover H$\alpha$, we assumed the ratio (H$\alpha$/H$\beta$) = 2.86 ratio to calculate H$\alpha$ luminosity from H$\beta$. 

{\vrc Given the  uncertainties in the absolute flux calibration of the 2dF spectra (see Sec.~\ref{sec-spec}), we calculated the H$\beta$ line luminosities from their equivalent widths and continuum levels estimated from the galaxy photometry.} We estimated the continuum level of each spectrum near H$\beta$ from the $g$ band absolute AB magnitude. We then multiplied the continuum level by the H$\beta$ equivalent width (calculated as part of the emission line fitting described above) to obtain the H$\beta$ luminosity. We corrected the line luminosity for dust using the \citet{Cardelli1989} dust extinction law as described above.

\section{Results: properties of WiggleZ Galaxies}
\label{sec-results}

\begin{figure*}
\includegraphics[width=\figwidth\linewidth]{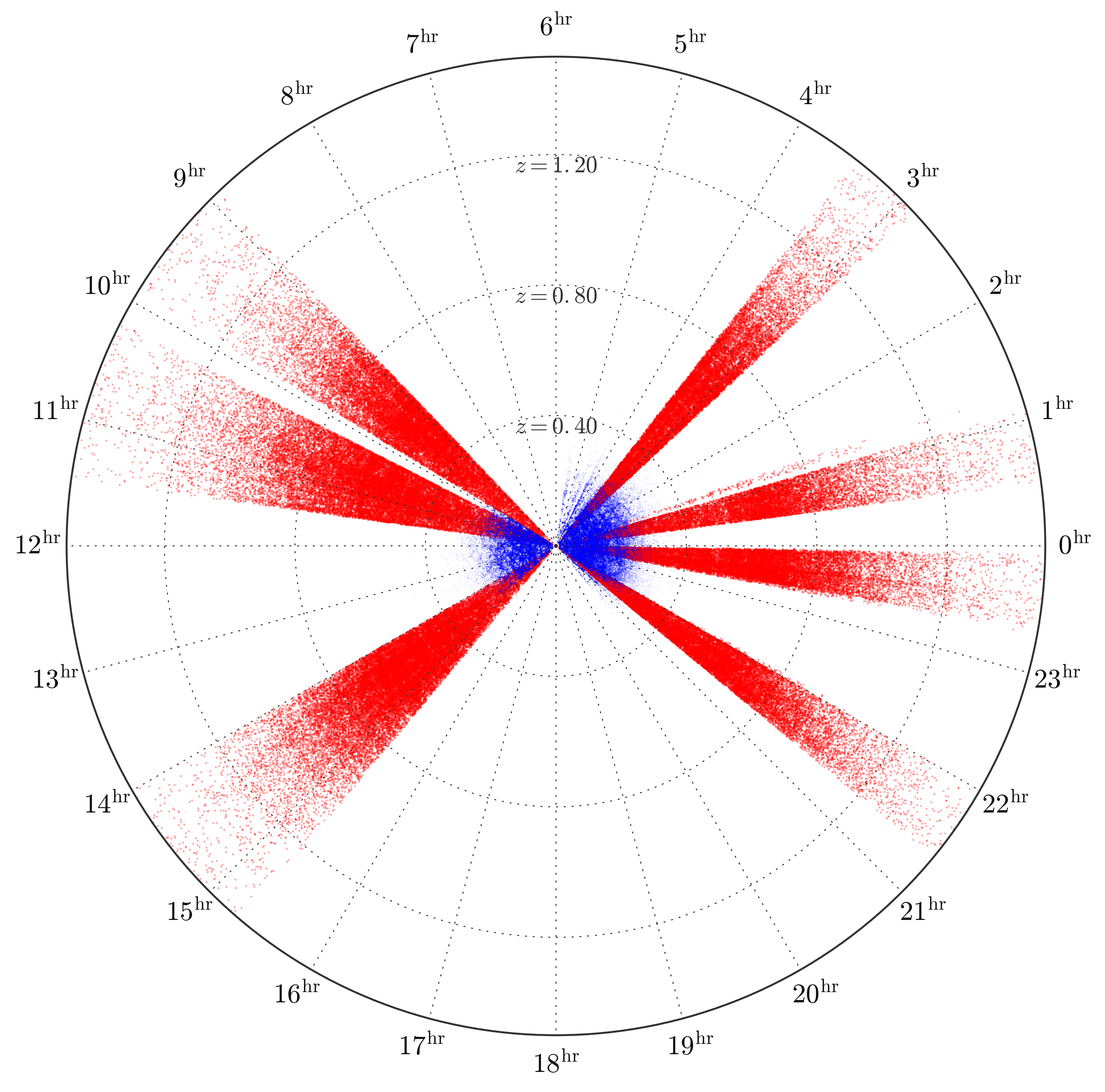}
\caption{A cone plot comparing the WiggleZ survey (red) to the low-redshift 2dFGRS survey \citep{Colless2001} in blue. Each point represents a galaxy with a secure redshift measurement, and we condense the 3D nature of the survey into a 2D representation by plotting only the right ascension as the angular coordinate. {\deltabf The redshift distribution for the WiggleZ survey can be seen in the figure, with consistently high observational density out to a redshift of approximately $z=0.9$.}
}
\label{fig-cone}
\end{figure*}

The final WiggleZ catalogue presented here consists of \nwig\ unique objects with reliable redshift measurements ($Q\ge 3$), which we plot on the sky in Fig.~\ref{fig-cone}. In this section we present some global analysis of this large data set. In all cases (unless noted otherwise) we restrict this analysis to the \nwigw\ galaxies selected for WiggleZ samples (including early observations with slightly different selection criteria; see Table~\ref{tab-targetcodes}). We do not include any spare fibre targets.

\subsection{Two-component Emission Lines}
\label{sec-width}

{\deltabf When we fitted the [OII], H$\beta$ and [OIII] lines, we tested each for a second broad component. The broad components of H$\beta$ (and possibly [OIII]) are correlated with AGN activity so we applied the analysis to spectra stacked according to FUV luminosity as described in Table~\ref{tab-2fits}. The table also lists the numbers of spectra we visually flagged (see Sec.~\ref{sec-spec}) as `AGN' in each luminosity bin: the `AGN' fraction increases rapidly at high luminosity. We discuss the results for each of the lines in turn.}

\begin{table*}
{\deltabfz
\caption{Two-component fits to emission lines in spectra averaged by FUV luminosity.}
\label{tab-2fits}\begin{tabular}{crcrrcccrccrr}
\hline
$M_{\FUV}$ & $N$  &  $\langle z \rangle$ & $N_{\rm AGN}$  & $f_{\rm AGN}$  & [OII]? &$F_{372.9}/F_{372.6}$& [OIII]? & {\vrc FW$_B$} &$F_{B}/F_{N}$& H$\beta$?& {\vrc FW$_B$} &$F_{B}/F_{N}$  \\
(mag) &   &  &  &  (\%) &  & &  & (\kms) && & (\kms)  &  \\
(1) & (2) & (3) & (4) & (5) & (6) & (7) & (8) & (9) & (10) & (11) & (12)& (13) \\
\hline
-25 -22 &    5557   &  1.43 &  3399 & (61.2)  &   n  &       -  &  Y  & {\vrc  800} & 0.95 &         Y  & {\vrc 3070} & 4.4    \\
-22 -21 &   16344  &  0.94 &    869 & (5.32)  &   n  &       -  &  Y  & {\vrc  610}& 0.51 &         Y  & {\vrc 940} &0.30  \\
-21 -20 &   64250  &  0.73 &    312  &(0.49)  &   Y  & 1.42  &  Y  &{\vrc 1010}& 0.16 &         n  & - & -   \\
-20 -19 &   72471  &  0.55 &      69 & (0.10)  &   Y  & 1.44  &  Y  &{\vrc 1030}& 0.10 & {\vrc n} & -  & $-$0.04  \\
-19 -18 &   32880  &  0.36 &        9 & (0.03)  &   Y  & 1.47  &  Y  &{\vrc   910}& 0.07 & {\vrc n}  & -  & 0.09 \\
-18 -17 &   13823  &  0.24 &        3 & (0.02)  &   Y  & 1.56  &  Y  &{\vrc   940}& 0.05 & {\vrc n} & -  & 0.16 \\
-17 -14 &    8817   &  0.13 &        7 & (0.08)  &   Y  & 1.45  &  n  &       - &   -     & {\vrc n}  & - & 0.08 \\
\hline
\end{tabular}

\parbox[t]{\textwidth}{
Notes. Each row of the table refers to a luminosity bin defined by the FUV absolute magnitude range in Column 1. The numbers of spectra and their average redshifts are given in Columns 2 and 3. Columns 4 and 5 give the number and fraction of the spectra manually flagged as AGN. {\vrd Columns 6, 8 and 11} indicate if a two component fit is not/very strongly preferred (corresponding to n/Y) according to the Bayes information criterion (see Sec.~\ref{sec-stackfits}). For [OII] this shows if the doublet is resolved, whereas for [OIII] and H$\beta$ it shows if there is a second, broad component. Column 7 gives the flux ratios of the [OII] doublet components. {\vrc Columns 9 and 12 give the full width at half-maximum of the broad components of [OIII] and H$\beta$ respectively.}  Columns 10 and 13 give the flux ratios of the broad and narrow components of [OIII] and H$\beta$ respectively.
} }
\end{table*}


We tested if the [OII] doublet (air wavelengths 372.60 and 372.88 nm) was resolved by adding a second line whose only free parameter was its flux (its wavelength and width being fixed with respect to the first line). We found that a 2-component fit was preferred (i.e.\  the doublet was resolved) for luminosities in the range $-21<M_{\FUV}<-17$. Where we resolved the [OII] doublet, we measured the flux ratio ($F_{372.9}/F_{372.6}$ in Table~\ref{tab-2fits}) which is a measure of electron density. The ratio decreases slightly with luminosity, suggesting that the emission regions in the more luminous galaxies are denser. These values all correspond to the limit of low electron density, $n_e<50 \cm^{-3}$ for $T=10^4$K \citep[see][Fig.~5.3]{Osterbrock1989}. 

\begin{figure}
\includegraphics[width=\figwidth\linewidth]{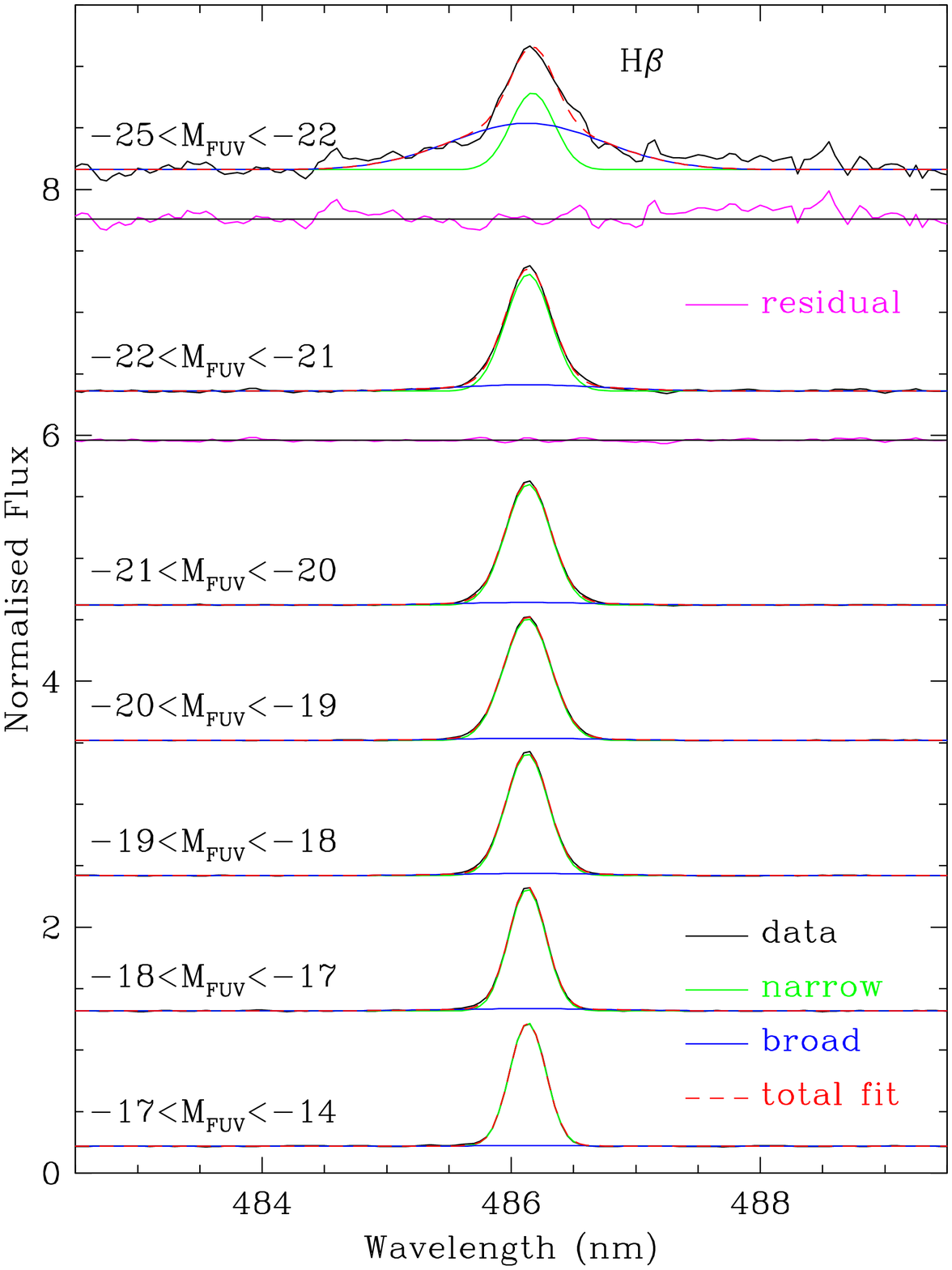}
\caption{Two-component fits to H$\beta$ emission lines in stacked spectra, averaged as a function of absolute FUV magnitude. The spectra are shown in black, the narrow components in green, the broad components in blue and the combined fit as a dashed red line. {\vrc The residuals (data $-$ total fit) are plotted in magenta against a zero line below the fits to the two most luminous samples.}
}
\label{fig-lineHb}
\end{figure}

{\deltabf We tested for a broad H$\beta$ component by adding a second line with wavelength, flux and width all as free parameters. The spectra and the fits are shown in Fig.~\ref{fig-lineHb}.} We detect  broad H$\beta$ components in our two most luminous samples, as expected for AGN (see Table~\ref{tab-2fits}). In the most luminous sample, the broad component has a FWHM line width of {\vrc 3070$\pm$260} \kms and contains 4.4 times more flux than the narrow component. The broad component of the second most luminous sample is less prominent with a FWHM line width of  {\vrc 940$\pm$120} \kms and only 0.30 times the flux of the narrow component. Unlike the broad wings of the [OIII] lines discussed below, these are not blue shifted compared to the line cores. {\deltabf The details of the fits are listed in Table~\ref{tab-2fits}}. 

\begin{figure}
\includegraphics[width=\figwidth\linewidth]{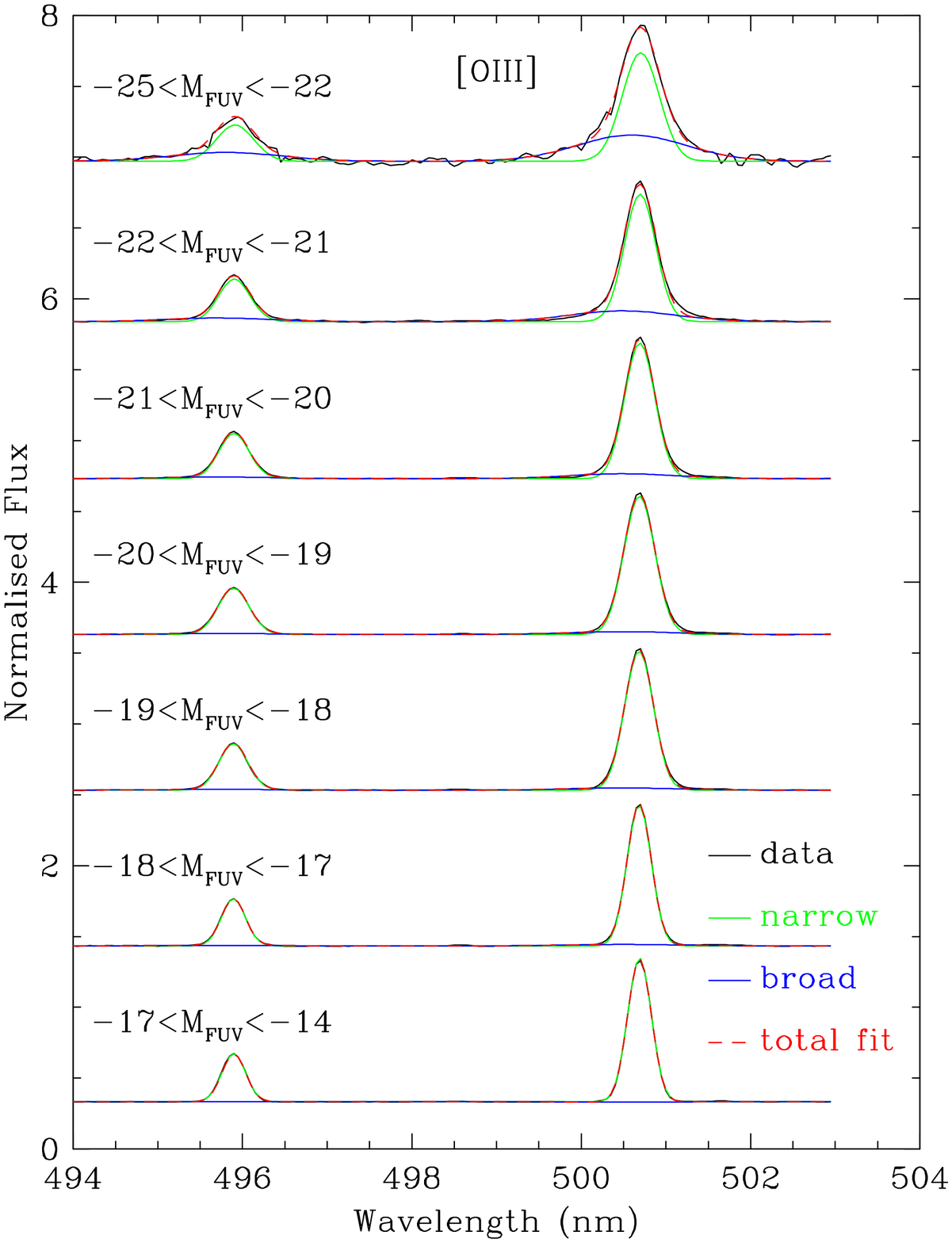}
\caption{Two-component fits to [OIII] emission lines in stacked spectra, averaged as a function of absolute FUV magnitude. The spectra are shown in black, the narrow components in green, the broad components in blue, and the combined fit as a dashed red line.
}
\label{fig-lineO3}
\end{figure}

{\deltabf We tested for a second component to the [OIII] doublet by adding a second doublet to the fit. {\vrc The wavelength, flux and width of the second doublet were free parameters, but the separation (495.9, 500.7 nm) was fixed.} The spectra and the fits are shown in Fig.~\ref{fig-lineO3}.} The [OIII] doublet lines have a significant additional component in the more luminous samples.  The extra components are broad and blue shifted compared to the narrow line components. {\deltabfz In the most luminous sample the flux of the broad component has 60 per cent of the flux of the narrow component, with a} FWHM width of 857 \kms, and a blueshift of 0.166 nm (100 \kms). 

\subsection{Mass Metallicity Relation}
\label{sec-massmetal}

\begin{figure}
{\deltabf
\includegraphics[width=\figwidth\linewidth]{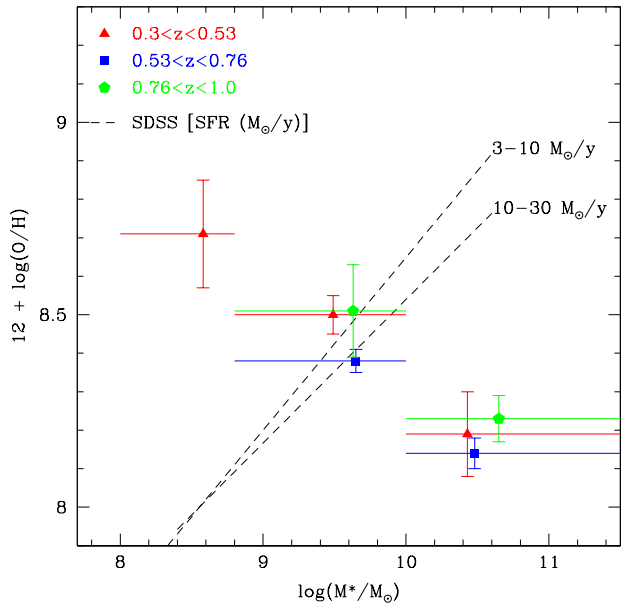}
\caption{The metallicities of WiggleZ galaxies compared to SDSS galaxies as a function of stellar mass. The WiggleZ values (points) are calculated for spectra averaged in bins of stellar mass and redshift. The range of mass for each point is indicated by the horizontal lines (see Table~\ref{table:massbins}). The dashed lines show trends of metallicity against stellar mass for different star formation rates in low-redshift SDSS galaxies \citep{Andrews2013}. The WiggleZ galaxies in the highest mass bins have significantly lower metallicities than given by the low-redshift SDSS relations for similar star formation rates.
}
\label{fig-metal}
}
\end{figure}

\begin{figure}
{\deltabf
\includegraphics[width=\figwidth\linewidth]{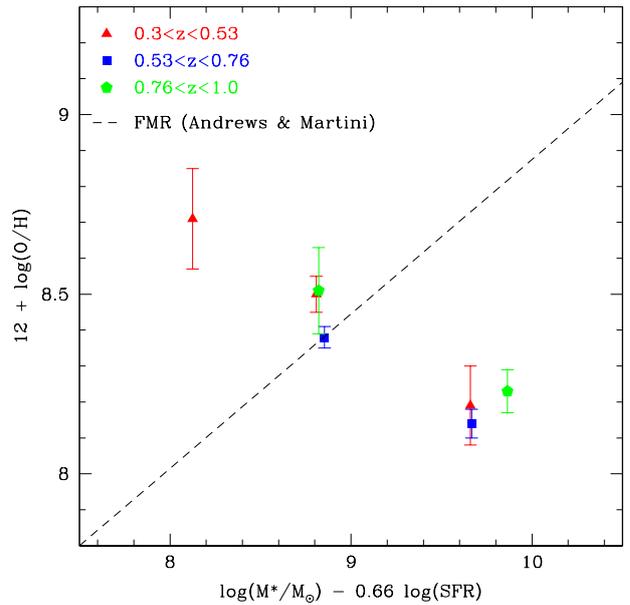}
\caption{The metallicities of WiggleZ galaxies projected on the fundamental metallicity relation (FMR) for SDSS galaxies. The WiggleZ values (points) are calculated for spectra averaged in bins of stellar mass and redshift (see Table~\ref{table:massbins}). The FMR predicts metallicity as a function of stellar mass and star formation rate \citep{Mannucci2010}; the projection used here was calibrated for direct-method metallicities by \citet{Andrews2013}. The dashed line shows the FMR for low-redshift SDSS galaxies \citep{Andrews2013}. The WiggleZ galaxies in the highest mass bins have significantly lower metallicities than predicted by the fundamental metallicity relation.
}
\label{fig-metalFMR}
}
\end{figure}

\begin{figure}
{\deltabf
\includegraphics[width=\figwidth\linewidth]{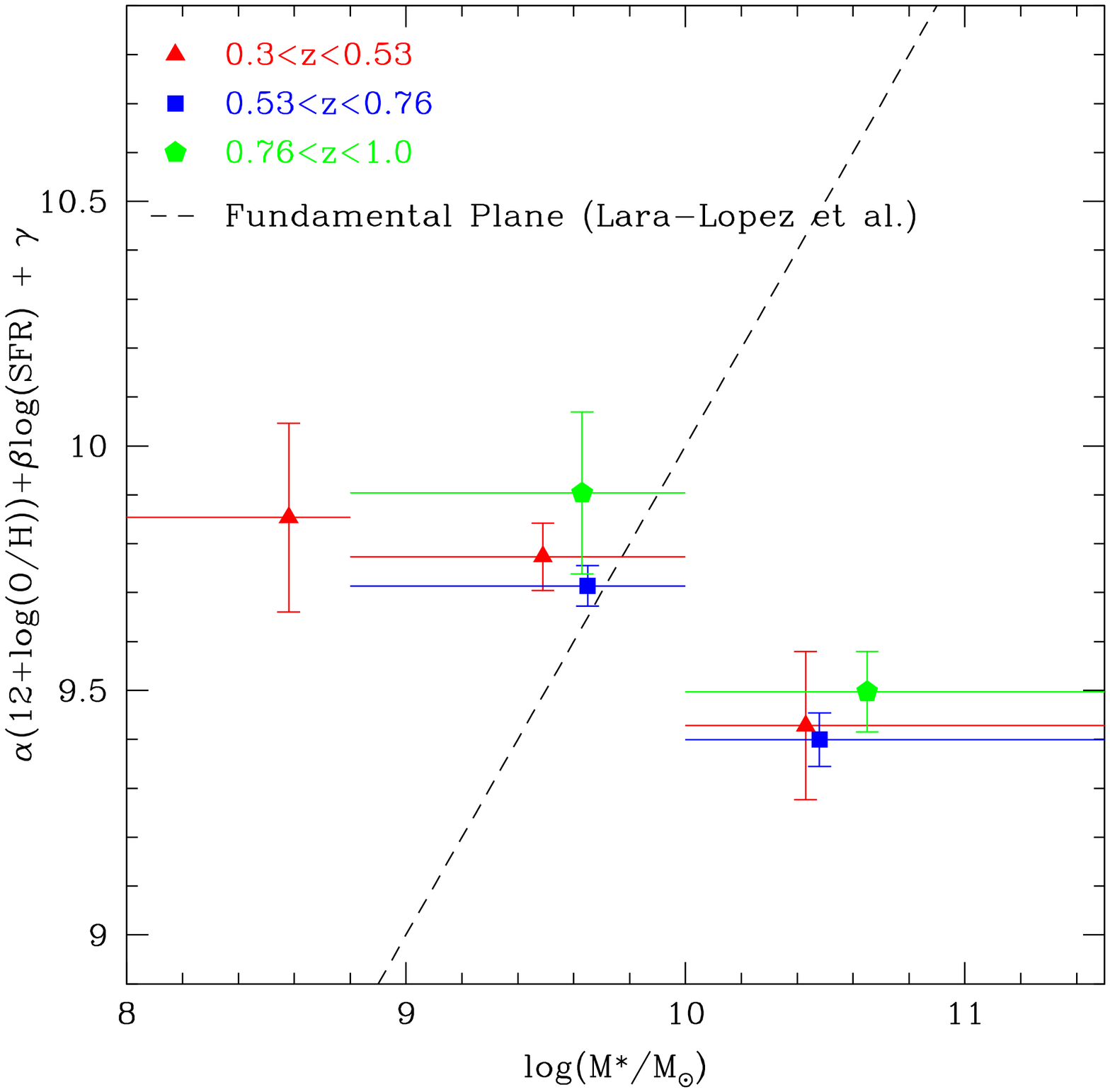}
\caption{WiggleZ galaxies projected on the \citet{LaraLopez2013} fundamental plane (FP). The stellar mass predicted by the FP (as in Equation 2) is plotted as a function of observed stellar mass. The WiggleZ values (points) are calculated for spectra averaged in bins of stellar mass and redshift. The range of mass for each point is indicated by the horizontal lines (see Table~\ref{table:massbins}). The dashed line shows the FP for low-redshift galaxies \citep{LaraLopez2013}. The FP relation predicts significantly lower masses than observed for WiggleZ galaxies in the highest mass bins.
}
\label{fig-metalFP}
}
\end{figure}

{\deltabf 
We present our metallicity measurements for the stacked WiggleZ spectra in Table \ref{table:massbins} and Figs.~\ref{fig-metal}, \ref{fig-metalFMR} and \ref{fig-metalFP}. We calculated average spectra in three bins of mass and three bins of redshift, but there were only sufficient galaxies for analysis in the low mass bin ($8.0<\log({M_*}/{M_\odot})<8.8$) at the lowest redshift. The table shows that, as the average masses of the WiggleZ spectra increase, so do their FUV luminosities. This is a consequence of the flux-limited selection of the WiggleZ sample. 

We show the mass-metallicity relation of the WiggleZ sample in Fig.~\ref{fig-metal} compared to the trends fitted by \citet{Andrews2013} to their low-redshift SDSS sample for different star formation rates. The WiggleZ metallicities are consistent with those of \citet{Andrews2013} at intermediate stellar masses ($8.8<\log({M_*}/{M_\odot})<10$), but fall above the \citet{Andrews2013} relations at low stellar mass and below at high stellar mass. At each mass where there are multiple measurements, there is no significant difference in the metallicity of the WiggleZ galaxies as a function of redshift.


\citet{Mannucci2010} have shown that the dependence of the mass-metallicity relation on star formation rate can be described by a  `fundamental metallicity relation' (FMR). The mass-metallicity-star formation rate relation becomes a single curve if it is projected into the plane defined by metallicity and a new parameter $\mu = \log(M_*/M_\odot) - \alpha \log(SFR)$ where the star formation rate $SFR$ is measured in units of $M_\odot/yr$. We plot the WiggleZ stacked spectra on the FMR in Fig.~\ref{fig-metalFMR} using the value of $\alpha=0.66$ calibrated for direct method metallicity measurements by \citet{Andrews2013}. This demonstrates a similar behaviour to Fig.~\ref{fig-metal} with the WiggleZ galaxies falling above the relation at low mass and below it at high mass.



An alternative projection of the mass-metallicity-star formation rate relation was proposed by \citet{LaraLopez2013}. This `fundamental plane' predicts the mass from metallicity and star formation rate with a relation of the form 
\begin{equation}
\log(M_*/M)=\alpha(12+\log(O/H))+\beta \log(SFR) + \gamma,
\end{equation}
with $\alpha = 1.376, \beta = 0.607,$ and  $\gamma = -2.550$. We plot this quantity for the WiggleZ stacked spectra in Fig.~\ref{fig-metalFP}. It demonstrates the same behaviour as the other projections with the WiggleZ galaxies falling above the relation at low mass, below it at high mass and not showing any significant change with redshift.
}

\begin{table*}
{\vrc
\caption{Line ratios of WiggleZ galaxies binned by stellar mass and redshift.}
\label{table:lineratios}
\begin{tabular}{cccrrccc}
\hline
$\langle \log({M_*}) \rangle$ (range) &  $\langle z \rangle$ (range) &  $\langle M_{\FUV} \rangle $& ${\vrd f_{MEx}}$  & $N$  &
$ {\rm [OII]} \over \rm H\beta $ & ${\rm [OIII](436.3)} \over \rm H\beta$ & ${\rm [OIII](500.7,495.9)} \over \rm H\beta$ \\
$(M_\odot)$ &  &(mag)& {\vrd (\%)} & &  & & \\
\hline
   8.58     (8.0--8.8)   &  0.40 (0.30--0.53) & -18.8  &  {\vrd 0.4} &   3158 & 1.809 &0.015 & 1.089   \\
   9.49    (8.8--10.0)  &  0.42 (0.30--0.53) & -19.0  &  {\vrd 0.7} & 28890 & 1.903 & 0.016 & 0.787   \\
   9.65    (8.8--10.0)  &  0.63 (0.53--0.76) & -19.9  &  {\vrd 0.9} & 23156 & 1.669 & 0.021 & 0.943   \\
   9.63    (8.8--10.0)  &  0.85 (0.76--1.00) & -20.7  &  {\vrd 0.4} &   5837 & 2.361 & 0.023 & 1.050   \\    
 10.43  (10.0--12.0)  &  0.44 (0.30--0.53) & -19.2  &  {\vrd 7.5} & 25070 & 1.517  & 0.010 & 0.375   \\
10.48  (10.0--12.0)   &  0.64 (0.53--0.76) & -20.0  &  {\vrd 12.8} & 48882 & 1.501  & 0.015 & 0.525   \\
10.65  (10.0--12.0)   &  0.86 (0.76--1.00) & -20.7  &  {\vrd 8.7} & 23343 & 2.424  & 0.024 & 0.741   \\    
\hline
\end{tabular}
\parbox[t]{\textwidth}{
Note: Each row describes one stacked spectrum. The first three columns give the average properties of the $N$ spectra used for each stack. {\vrd The fourth column, $f_{MEx}$, gives the percentage of spectra rejected by the MEx criterion.} The final three columns give line ratios measured from the stacked spectrum.
}
}
\end{table*}
 
\begin{table*}
{\vrc
\caption{Derived properties, including metallicity, of WiggleZ galaxies binned by stellar mass and redshift.}
\label{table:massbins}
\begin{tabular}{ccrrrrrrc}
\hline
$\langle \log({M_*}) \rangle$ (range) &  $\langle z \rangle$ (range) & $N$ & $E(B-V)$ & $T_e$[OIII] & $O^+/H^+$ & $O^{++}/H^+$ &  $ SFR $ &  $12+\log(O/H)$    \\
$(M_\odot)$ &  & & (mag) & (K) & & & $(M_\odot/{\rm yr})$  &  (dex) \\
\hline
  8.58     (8.0--8.8)   &  0.40 (0.30--0.53) &   3158 & 0.438 & 9325  & $3.7\times 10^{-4}$ & $1.7\times 10^{-4}$ & 4.87   & $8.71 \pm 0.14$\\
   9.49    (8.8--10.0)  &  0.42 (0.30--0.53) & 28890 & 0.537 & 10686 & $2.4\times 10^{-4}$ & $7.2\times 10^{-5}$ &  10.8   & $8.50 \pm 0.05$\\
   9.65    (8.8--10.0)  &  0.63 (0.53--0.76) & 23156 & 0.268 & 10558 & $1.5\times 10^{-4}$ & $9.1\times 10^{-5}$ &  16.1  & $8.38 \pm 0.03$\\
   9.63    (8.8--10.0)  &  0.85 (0.76--1.00) & 5837 &   -0.080 & 9707 & $1.9\times 10^{-4}$ & $1.5\times 10^{-4}$ & 16.8   & $8.51 \pm 0.12$\\   
 10.43  (10.0--12.0)  &  0.44 (0.30--0.53) & 25070 &  0.509 & 11778 & $1.4\times 10^{-4}$ & $2.6\times 10^{-5}$ &  14.7   & $8.19 \pm 0.11$\\
10.48  (10.0--12.0)  &  0.64 (0.53--0.76) & 48882 &   0.291 & 11575 & $1.0\times 10^{-4}$ & $3.8\times 10^{-5}$ & 17.1   & $8.14 \pm 0.04$\\
10.65  (10.0--12.0)  &  0.85 (0.76--1.00) & 23343 &   -0.037 & 11286 & $1.1\times 10^{-4}$ & $6.0\times 10^{-5}$ & 15.5   & $8.23 \pm 0.06$\\
\hline
\end{tabular}
\parbox[t]{\textwidth}{
Note: Each row describes one stacked spectrum. The first two columns give the average properties of the $N$ spectra used for each stack. The final six columns give properties derived from the line ratios of the stacked spectrum.
} }
\end{table*}

\section{Discussion}
\label{sec-discuss}

\subsection{Resolving the [OII] doublet.}

We expect the [OII] doublet to be unresolved at low luminosities as these correspond to low redshifts, but it should be increasingly better resolved at higher luminosities in our sample due to the increasing redshift\footnote{At a redshift of zero the 0.28 nm separation of the [OII] doublet is less than the 0.36 nm resolution of the spectrograph (see Sec.~\ref{sec-spec}); at our median redshift of 0.6 the separation is 0.45 nm.}. However, it is not resolved in our two most luminous bins, presumably because the intrinsic line width due to internal broadening in these luminous galaxies has merged the two doublet lines. For comparison, the [OIII] line widths are 383 \kms in the most luminous bin, larger than the 225 \kms separation of the [OII] doublet. This issue of the [OII] line being blended was discussed as a possible limitation of single-line redshifts in the DEEP2 survey (0.14 nm resolution so the doublet should always be resolved) by \citet{Kirby2007}: we have now identified that UV luminosity is a strong predictor of this blending. 

The value of the [OII] doublet ratio was discussed by \citet{Comparat2013} in the context of designing galaxy surveys that would rely solely on detection of a resolved doublet to measure redshifts. Their models assumed a lower canonical ratio ($F_{372.9}/F_{372.6}=1$) than we measure for the WiggleZ galaxies, but this is not likely to make a large difference to the detectability of the doublet. We confirm the trends predicted by \citet{Comparat2013} that detection of the doublet increases with redshift and luminosity with the exception of our highest luminosity ranges.

\subsection{Active galaxies in the WiggleZ sample}
\label{sec-fAGN}

The broad blue shifted component we detect in the [OIII] lines is consistent with the wings measured in SDSS QSOs \citep[][median offsets of 162 and 97 \kms for Type 1 and 2 AGN]{Peng2014}. They report median velocity dispersions of 393 and 370 \kms for Type 1 and 2 AGN. These correspond to FWHM widths of 925 and 871  \kms respectively, similar to the WiggleZ value of 857 \kms. While \citet{Peng2014} find similar fluxes in the wing and core components of the lines, we find that the wings only contain 60 per cent of the flux of the line cores in our average spectra. 
\citet{Peng2014} discuss the blue wings in terms of gas outflows driven either by the central AGN or by intense star formation activity. They argue that since the outflow velocity in their sample did not correlate with the AGN properties, there must be a significant contribution from star formation, although we suggest that the AGN variability might mask any correlations. High-resolution resolved spectroscopy would be a possible way to separate these mechanisms as \citet{Peng2014} indicate, but this would be challenging for a large sample of WiggleZ galaxies due to their high redshifts, as well as the complex structures revealed in the small sample already measured by \citet{Wisnioski2011}.


The broad H$\beta$ component we detect in the more luminous galaxies is a clear indication of AGN in the sample. \citet{Wisnioski2012} also report a broad H$\beta$ component in the stacked spectra of eight individual star forming clumps found in three high-redshift ($z \sim 3$) WiggleZ galaxies, but this was narrower (FWHM 490 \kms) and not associated with AGN activity as the clumps were not central in their host galaxies.

{\deltabf Our detection of the broad H$\beta$ component is correlated with the fraction of spectra manually flagged as AGN in these luminosity bins (also shown in Table~\ref{tab-2fits}). In the highest luminosity sample ($-25<M_{\FUV}<-22$), 61 per cent of the galaxies were manually flagged as AGN. \citet{Jurek2013} found evidence that the WiggleZ sample contained additional AGN at these luminosities which had not been manually identified. We inspected the individual spectra of the high luminosity galaxies which had not been marked as AGN and some of them certainly do display broad lines, but the quality of individual spectra is too low in many cases to classify them as AGN or not. This means we cannot categorically confirm if an individual WiggleZ galaxy is an AGN, but for the purpose of any future analysis, the FUV luminosity is the best proxy for AGN classification, with substantial (more than 60 per cent) AGN contamination present in galaxies more luminous than $M_{\FUV}=-22$ mag. There is at least 5 per cent AGN contamination in the next lower luminosity range, ($-22<M_{\FUV}<-21$),} {\vrc with both contamination rates estimated from the number of spectra manually flagged as AGN.

The flux-limited nature of the WiggleZ sample means that the UV-luminous galaxies with high AGN rates are also at the highest redshifts in the sample (see Table~\ref{tab-2fits}). Many fewer galaxies were flagged as AGN at more typical redshifts for the survey ($0.2<z<1.0$); at these redshifts we can use the mass-excitation diagnostic to identify probable AGN as described in Sec.~\ref{sec-mex}.  
}


\subsection{The metallicity of high-mass star forming galaxies}
\label{sec-dis-metal}


{\deltabf The main features of our direct method metallicity measurements in Table~\ref{table:massbins} and Figures~\ref{fig-metal} to \ref{fig-metalFP} is that the metallicities of the WiggleZ galaxies fall significantly below those of normal emission-line galaxies at high stellar masses and they do not appear to exhibit any significant change with redshift. Our one low-mass measurement has a higher metallicity than normal galaxies.}

We compare the WiggleZ metallicities to the mass-metallicity relations of local ($z<0.1$) star-forming galaxies \citep{Andrews2013} in Fig.~\ref{fig-metal}. The WiggleZ metallicities are consistent with those of the local galaxies at intermediate stellar mass, but they fall significantly below the local relations at stellar masses greater than $10^{10} M_\odot$. There is a possible measurement bias between the two samples as the data measure different physical regions of the respective galaxies. The WiggleZ data use 2 arc second-diameter fibres. In average seeing of 2 arc seconds, this corresponds to projected radii of 8.5 kpc at $z=0.5$ and 10.0 kpc at $z=0.7$. By comparison, the 3 arcsec SDSS fibres in 1.5 arcsec seeing sample projected radii of 3.1 kpc at $z=0.078$. Thus the SDSS data preferentially sample the inner regions of galaxies which could be expected to have higher-than-average metallicities. \citet{Tremonti2004} found this effect is SDSS galaxies: the metallicities varied by up to 0.1 dex with redshift in the SDSS sample but they were not able to determine an absolute correction. \citet{Tissera2015} measured the metallicity gradients of simulated star forming disc galaxies: the average metallicity change between radii of 2 kpc and 6 kpc is between 0.1 and 0.25 dex. This is unlikely to lead to a significant bias compared to the uncertainty in the WiggleZ metallicities, especially considering that the WiggleZ values are consistent with the low redshift relations in our three lower mass bins which span a redshift range of $0.5<z<0.6$ (Table~\ref{table:massbins}). We use the same direct method for metallicity measurement on average spectra as \citet{Andrews2013} so this difference is unlikely to be due to our calibration. 

{\deltabf 
As we discuss in Sec.~\ref{sec-intro}, some authors report evolution of the mass-metallicity relation with lower metallicities at higher redshift, although many authors fail to detect any evolution. We considered possible evolution because the WiggleZ galaxies have significantly higher redshifts ($z=0.5-0.7$) than the SDSS galaxies \citep[at a median redshift of $z=0.078$,][]{Andrews2013} -- a difference in look-back time of about 5 Gyr. However, inspection of Fig.~\ref{fig-metal} reveals that the WiggleZ metallicity values do not vary with redshift at a given stellar mass. Furthermore, at our middle mass range ($8.8<\log({M_*}/{M_\odot})<10$) the WiggleZ metallicities are entirely consistent with the low-redshift \citep{Andrews2013} sample at all redshifts. We therefore conclude that the low metallicities of the high-mass WiggleZ galaxies are not due to evolution with redshift.

A further possible explanation of the low metallicities of the high-mass WiggleZ galaxies might be increased contamination of the sample by AGN at higher masses. Shocks in some AGN can increase the measured electron temperature \citep[e.g.][]{dopita1995}; this in turn will lead to a reduced estimate of the metallicity. We tested for the effect of AGN contamination by repeating the measurements in Table~\ref{table:massbins} on stacked spectra where we excluded the most $FUV$-luminous galaxies -- those with $M_{FUV}<-22$. Galaxies in this range are over 60 per cent AGN as we show above. The results were identical (within uncertainties) after removing the likely AGN, so we conclude that the low metallicities are not due to AGN contamination.

The fundamental difference between the WiggleZ galaxies and those galaxies used to define all the comparison relations we show in Figures~\ref{fig-metal} to \ref{fig-metalFP} is the extreme UV selection of WiggleZ galaxies. The WiggleZ galaxies were selected by their UV fluxes and colours (designed to detect the Lyman break at redshifts around $z=0.5$; see Sec.~\ref{sec-selection}). Our successful selection of Lyman break galaxies was confirmed by the UV-optical colours and the clustering of the WiggleZ galaxies (see Paper 1). These are extreme galaxies, having median star formation rates (which strongly correlate with UV luminosity) in the top 5 per cent of optically-selected galaxies \citep{Jurek2013}.  The metallicities of the extreme WiggleZ galaxies fall significantly below those of normal galaxies at stellar masses greater than around $10^{10} {M_\odot}$.

There have been several previous reports of subsets of galaxies which have unusually low metallicities for their mass or luminosity. These have been attributed to starburst activity \citep[e.g.][]{Hunt2012}, particularly that driven by galaxy mergers \citep[e.g.][]{Lee2004}. \citet{Hunt2012} use the term `low metallicity starburst' to describe galaxies with this behaviour. This category included galaxies ranging from local blue compact dwarf (BCD) galaxies and luminous compact galaxies (at $z\approx 0.3$) to Lyman break galaxies (at $z>1$). The most extreme examples (the BCDs) show a trend that is qualitatively similar to the WiggleZ galaxies in that their metallicities as much as 1 dex lower than normal at masses of around $10^{10} {M_\odot}$, but they lie on the normal relation at lower masses \citep[{\vrc see fig.\ 3 of}][]{Hunt2012}.

Both \citet{Mannucci2010} and \citet{Troncoso2014} measured samples of high ($z\approx 3$) redshift galaxies that fall significantly below their fundamental metallicity relation (FMR; we show the relation recalibrated for direct metallicities in Fig.~\ref{fig-metalFMR}). They used the Lyman break method at optical wavelengths to select high-redshift star-forming galaxies. \citet{Troncoso2014} discuss four possible explanations of the low metallicities of these galaxies. They tested for merger activity, but found this was not correlated with the deviations from the FMR. They considered that the (blue) colour selection would bias the sample against metal-rich dusty galaxies, but discounted this explanation as the low redshift sample also included colour-selected galaxies and the deviation was greatest for the most massive (and therefore most dusty) galaxies, concluding that the deviation was specific to high-redshift galaxies. They discussed a possible transition in galaxy evolution at redshifts in the range $2.5<z<3$, their preferred explanation. Finally they considered that Lyman break galaxies may form a class that do not follow the FMR at any redshift, noting the need for independent samples to test this possibility. The WiggleZ galaxies do represent an independent sample of Lyman break galaxies, but their selection makes them much more extreme compared to normal galaxies \citep[see][]{Jurek2013} so we cannot compare them directly to the \citet{Mannucci2010} and \citet{Troncoso2014} galaxies. {\vrc In addition, we should note that \citet{Mannucci2010} and \citet{Troncoso2014} both used strong-line metallicity diagnostics which may not be reliable at high redshift \citep{Ly2016b}.}

In summary, the WiggleZ galaxies are not the only galaxies reported to have low metallicities for their stellar masses at masses greater than around $10^{10} {M_\odot}$. The types of galaxy reported seem to vary considerably, from blue compact dwarf galaxies \citep{Hunt2012} to Lyman break {\vrc  galaxies \citep{Mannucci2010,Troncoso2014}.} \citet{Hunt2012} suggest that the low metallicities are associated with extreme starburst activity: this is certainly true of the WiggleZ galaxies, but may not be true of the high-redshift Lyman break galaxies. 
}



\section{Final Public Data Release}
\label{sec-data}

The final WiggleZ Dark Energy Survey catalogue contains data for \nwig\ unique objects. We present the data as a catalogue as well as individual spectra. We describe these data products and their access in this section. 
\subsection{Tabular Data}
\label{sec-columns}

We present a sample of five lines of the catalogue in Table~\ref{tab-sample}; the rest of the catalogue is available online as described in Sec.~\ref{sec-online}. The catalogue has 30 data columns which we describe as follows.

\noindent{\bf Column 1: Name.} The name for each galaxy is based on its J2000 coordinates. The full IAU name  includes the survey name in the format `WiggleZ NhhJHHMMSSsss+DDMMSSss'. The first letter `N' describes the target selection category as in Table~\ref{tab-targetcodes}. This is followed by a two-digit code `hh' identifying the survey field by its Right Ascension in hours. This is followed by `J' to signify J2000 equinox coordinates which are encoded in the remaining digits HHMMSSsss and $\pm$DDMMSSss. For example, the name S11J101510859+04305665 refers to a SDSS target in the 11-hour field with J2000 coordinates 10:15:10.859, +04:30:56.65. 

\noindent{\bf Columns 2-3: Coordinates.}
The next two columns give the J2000 object coordinates in decimal degrees. These are taken from the SDSS or RCS2 catalogues depending on the survey region. {\deltabf Note that we identified small but significant offsets (listed in Table~\ref{tab-offset}) between the USNO-B reference frame used to define the RCS2 astrometry and a standard reference frame as measured by the 2MASS survey. We corrected the RCS2 positions for the offsets, so that all positions in the final catalogue are consistent with the 2MASS coordinates.}

\noindent{\bf Columns 4-6: Redshift value, uncertainty and quality.}
The measured redshift, its uncertainty and the quality code $Q$. The meanings of the quality code values are given in Table~\ref{tab-quality}.

\noindent{\bf Columns 7-13: Photometry.}
The final catalogue includes UV and optical photometry for all sources. The UV (FUV, NUV) measurements are from the {\it GALEX} satellite data and the optical data are from SDSS in the equatorial fields ($u g r i z$) and RCS2 in the Southern fields (only $g r z$). The photometry was all corrected for Galactic dust extinction as a function of the local E(B-V) reddening as described in Paper 1. 

\noindent{\bf Columns 14-20: Photometry uncertainty.}
The uncertainties for all the photometry are given in these columns.

\noindent{\bf Column 21: Reddening $E(B-V)$.} This column measures the local $E(B-V)$ reddening used to correct all the photometric data for Galactic dust extinction. 

\noindent{\bf Column 22: Survey class flag.} In addition to the short code at the start of each object name (Column 1) we provide a longer flag to specify the selection for each object. These are also listed in Table~\ref{tab-targetcodes}. There are two additional versions of the main survey flags, `WIG\_RCS2\_EXT' and `WIG\_SDSS\_EXT'. The `EXT' code means these objects form an `extended' sample: they were observed early in the survey, but did not satisfy the final selection criteria. This was a result of revisions to some early versions of the input catalogues.

\noindent{\bf Column 23: UT date.}
The Universal Time date when the associated spectrum was observed at the Anglo-Australian Telescope, in an integer format YYYYMMDD.

\noindent{\bf Column 24: FUV absolute magnitude.}
The absolute magnitude in the FUV band, calculated from the NUV observed-frame apparent magnitude, using median k-corrections \citep{Jurek2013} as a function of redshift.

\noindent{\bf Columns 25-26: Fitted properties.}
{\deltabf The stellar masses ($\log(M_*/M_\odot)$) and their uncertainties, calculated as by \citet{Banerji2013}; see Sec.~\ref{sec-masses}.}

\noindent{\bf Column 27: Spectrum filename.}
The name of the corresponding spectral data file. These are simply formed by numbering the files in sequence. The spectral files are described below.

\noindent{\bf Column 28: Spectrum comment.}
The final column gives the manual comment (if any) made by the person measuring the redshift of each galaxy. The most common observation here is `AGN' if the spectrum has AGN-like broad emission lines. A spectrum may also be flagged `fringed' or similar if it is badly affected by fringing (this is not represented by the $Q$ flag because a reliable multi-line redshift can sometimes be obtained despite very bad fringing).

\begin{table*}
\caption{The WiggleZ Catalogue. The full table is available online.}
\label{tab-sample}
\begin{tabular}{lrrrrrrrrrrrrrrrrrrrrrrrr}
\hline
           WiggleZ\_Name      &      RA     &      Dec  &redshift&  {\deltabfz $\Delta$redshift} & Q &   {\deltabfz $FUV$} &   {\deltabfz $NUV$} &   {\deltabfz $u$} &   {\deltabfz $g$} &  {\deltabfz $r$} &  {\deltabfz $i$ }  \\ 
                      1  &           2 &         3 &       4&       5 & 6 &     7 &     8 &     9  &   10   &  11    &    12   \\ 
\hline
 R01J003046635-02074486  &   7.6943130 & -2.129129 & 1.63248& 0.00170 & 3 &  null & 22.03 &   null &  21.23 &  21.01 &   null  \\ 
 R01J003046712+04223764  &   7.6946340 &  4.377121 & 0.75119& 0.00023 & 4 &  null & 22.69 &   null &  22.34 &  21.86 &   null  \\ 
 R01J003046734+01190408  &   7.6947255 &  1.317801 & 0.59666& 0.00003 & 3 & 24.05 & 22.58 &   null &  21.86 &  20.80 &   null  \\ 
 R01J003046795+04325062  &   7.6949790 &  4.547395 & 0.59860& 0.00019 & 4 & 24.08 & 22.20 &   null &  22.80 &  22.36 &   null  \\ 
 R01J003046812+00005664  &   7.6950495 &  0.015734 & 0.39627& 0.00015 & 4 & 23.46 & 22.28 &   null &  22.33 &  21.76 &   null  \\  
\hline
\end{tabular}
\vspace{3mm}

\begin{tabular}{rrrrrrrrrrrrrrrrrrrrrrrrrrrrr}
\hline
   {\deltabfz $z$ }& {\deltabfz $\Delta FUV$} & {\deltabfz $\Delta NUV$ }& {\vrc $\Delta u$} & {\deltabfz $\Delta g$ }& {\deltabfz $\Delta r$ } & {\deltabfz $\Delta i$ } &  {\deltabfz $\Delta z$} &   E(B-V)&     class &        {\deltabfz UT date} &   {\deltabfz $M_{FUV}$} &   {\deltabfz mass} & {\deltabfz $\Delta$mass }\\
   13   &    14 &    15  &    16  &    17  &     18 &     19 &     20 &       21&       22  &        23     &     24  &      25 &      26 \\
  \hline 
  20.79 &  null &   0.20 &   null &   0.02 &   0.01 &   null &   0.05 &   0.0302& WIG\_RCS2 &      20090924 &  -22.62 &    null &    null \\
  21.31 &  null &   0.20 &   null &   0.03 &   0.03 &   null &   0.08 &   0.0195& WIG\_RCS2 &      20090820 &  -20.12 &   10.32 &    0.11 \\
  19.75 &  0.34 &   0.17 &   null &   0.02 &   0.02 &   null &   0.03 &   0.0238& WIG\_RCS2 &      20090919 &  -19.66 &   10.87 &    0.16 \\
  22.65 &  0.28 &   0.17 &   null &   0.05 &   0.05 &   null &   0.20 &   0.0172& WIG\_RCS2 &      20090820 &  -20.05 &    8.84 &    0.25 \\
  21.40 &  0.24 &   0.14 &   null &   0.03 &   0.03 &   null &   0.09 &   0.0237& WIG\_RCS2 &      20080927 &  -18.98 &    9.30 &    0.29 \\   
\hline
\end{tabular}
\vspace{3mm}

\begin{tabular}{rrrrrrrrrrrrrrrrrrrrrrrrrrr}
\hline
            SpecFile  &   comments \\
                  27 &       28   \\
\hline
    wig000352.fits &   AGN      \\
    wig000353.fits &   -        \\
    wig000354.fits &   -        \\
    wig000355.fits &   -        \\
    wig000356.fits &   fringe   \\   
\hline
\end{tabular}

\end{table*}

\begin{table}
\caption{Object Name Codes.}
\label{tab-targetcodes}
\begin{tabular}{llrr}
\hline
Code, Flag &  Description & N & (\%)  \\
Col.\ 1, Col.\ 22 & & & \\
\hline
G, SPARE\_GALEX& UV variable sources & 67     & (0.03)  \\
R, WIG\_RCS2           & RCS2 regions (main) & 67956     & (30.1)   \\
R, WIG\_RCS2\_EXT & RCS2 regions (extra) & 21188     & (9.4)   \\
S, WIG\_SDSS           & SDSS regions (main) & 99475    & (44.1)    \\
S, WIG\_SDSS\_EXT & SDSS regions (extra) & 31692    & (14.1)    \\
X, SPARE\_RCS2 & RCS2 cluster galaxies & 2337 & (1.0)   \\
Y, SPARE\_RADIO & radio galaxies & 2700       & (1.2)  \\
Total      & & 225415                         & (100)  \\
\hline
\end{tabular}
\end{table}

\subsection{Spectral Data}

The spectral data accompanying the catalogue consist of a single FITS files for each object. Each file has two FITS extensions, the first (extension 0) contains the spectrum, and the second (extension 1) contains the variance. The header data includes extensive history data for the data processing but also the key parameters listed in Table~\ref{tab-keywords}.

The name of each spectrum file is given in Column 24 of the main catalogue. This allows software such as {\sc topcat} to automatically plot the spectrum corresponding to a catalogue entry. An important qualification about the spectral data is that a threshold of $\pm$30,000 counts has been applied to the data values. This means that any pixel with values outside this range is set to zero in the spectrum. This has been applied to remove major artefacts (such as large sky subtraction residuals or cosmic ray events) from the data processing so that most spectra can be plotted easily by routines that use automatic scaling to the total range of the data.

\begin{table}
\caption{{\vrc Main FITS header keywords in spectral files.}}
\label{tab-keywords}
\begin{tabular}{lllr}
\hline
Code &  Description  \\
\hline
OBJECT   &  WiggleZ target name \\
REFCODE  &  NASA ADS reference to this paper \\
RA\_OBJ  &  RA J2000 decimal degrees \\
DEC\_OBJ &  Dec J2000 decimal degrees \\
ZDSTART  &  Zenith distance at the start of exposure (deg) \\
EXPOSED  &  Exposure (sec) \\
UTSTART  &  UT time of exposure start \\
UTDATE   &  UT date of exposure start \\
DICHROIC &  Name of spectrograph dichroic \\
APERDIA  &  Diameter of fibre aperture \\
APUNIT   &  Units of fibre aperture {\deltabf diameter} (APERDIA) \\
\hline
\end{tabular}
\end{table}

\subsection{WiggleZ Data Access}
\label{sec-online}

We provide public online access to the WiggleZ data presented in this paper via two primary sources.

\subsubsection{Virtual Observatory Data Services}

The Australian National Computational Infrastructure service has developed and deployed a suite of Virtual Observatory {\deltabfz data services to access the WiggleZ data.} These services are part of the All-Sky Virtual Observatory (ASVO, \url{http://www.asvo.org.au/}) project. In accordance with the objectives of the ASVO, these services are compliant with the relevant services standards as defined by the International Virtual Observatory Alliance (IVOA, \url{http://www.ivoa.net/}). Compliance with these standards enables seamless access by tools such as the widely used {\sc topcat} tool (\url{http://www.star.bris.ac.uk/~mbt/topcat/}).

The primary service for accessing the WiggleZ data is a table access protocol (TAP) service, which allows a user to pose arbitrary SQL-like queries. This service is designed to be interacted with via compatible client tools, such as {\sc topcat}, as both the query strings and result sets which a TAP service generates are not human-readable. When driven by the {\sc topcat} tool, the TAP service provides the query results as a table which {\sc topcat} can visualise as required.

A simple ConeSearch (\url{http://www.ivoa.net/documents/latest/ConeSearch.html}) service has also been deployed, and can also be interacted with via a tool like {\sc topcat}. Results from the ConeSearch service include a link to the corresponding WiggleZ spectrum file (under the column `specfile'). Using a tool such as wget, or copying this link into a browser, the user is able to download the relevant WiggleZ spectrum file {\vrc directly to their computer.}

In order to locate the WiggleZ data services, and in line with IVOA best practices, the deployed services have been registered with one of the primary Virtual Observatory service registries, in this case operated by the EuroVO project (\url{http://www.euro-vo.org/}). The {\sc topcat} tool is fully interoperable with this service. The search is done by opening a service (such as table access protocol, TAP) from the `VO' tab in the {\sc topcat} main window. The keyword `wigglez' is entered and the `Find Services' command should locate the data service which can then be queried directly. If the default search settings do not work, we recommend the EuroVO registry (\url{http://registry.euro-vo.org/services/RegistrySearch} with the protocol set to ` RI1.0')  in preference to other services as it is more likely to be up-to-date. 
 
In this way, users with the {\sc topcat} tool installed on their desktop can locate and interrogate the relevant WiggleZ data services, query these services directly, then download and visualise the search results, all using a single integrated work-flow from the one desktop tool.

%

The EuroVO registry may also be searched directly to discover the WiggleZ data services and view the related metadata. A user wishing to utilise the data services directly (e.g., via a shell script or an alternative tool to {\sc topcat}) may obtain the current direct URLs for each service in this way. The search interface for this registry is currently at \url{http://registry.euro-vo.org}.


The National Computational Infrastructure has committed to hosting the WiggleZ data services in the near term as part of its ongoing commitment to the ASVO project, which is deploying IVOA-compliant services to a range of important astronomical data-sets.

\subsubsection{Online Catalogues}
We will publish the object catalogue and spectrum files at the Centre de Donn\'ees astronomiques de Strasbourg (CDS). The CDS data will be directly linked to this paper and are in the form of a plain-text catalogue file plus a compressed collection of the single spectrum files in FITS format.

We will also publish the full catalogue and spectra as data products associated with this paper on the NASA Extragalactic Database (NED).

\section{Summary}
\label{sec-summary}

To emphasise the large size of the WiggleZ Dark Energy Survey we note that we measured \nwig\ unique redshifts, of which 139,785 have redshifts $z>0.5$. At the time of writing Paper 1, the total number of $0.5<z<1.5$ redshifts published from all surveys was 100,000 (as listed by NED). The WiggleZ survey, now completed, has more than doubled this quantity.

In this paper we have presented the final data set of the WiggleZ survey as a publicly available catalogue with spectra of \nwig\ objects. In addition, we include stellar mass estimates for 82 per cent of the WiggleZ galaxies.

{\deltabf We have used average spectra stacked in bins of UV luminosity to measure various emission-line properties of the WiggleZ galaxies. The [OII] 372.7 nm doublet was resolved, except for the most luminous objects where it was confused by line broadening. We measured ratios of $F_{372.9}/F_{372.6}=1.4$ to 1.6 when the doublet was resolved. In the most luminous ($M_{FUV}<-21$) galaxies, both the [OIII] and H$\beta$ lines exhibited broad second components. The broad [OIII] components {\vrc (full width half-maximum 800 \kms)} were blue shifted by 100 \kms, indicative of gas outflows. The broad components of the H$\beta$ lines were extremely broad  {\vrc (full width half-maximum 3070 \kms), a clear indication} of AGN activity. The FUV luminosity is a strong indicator of AGN activity in the sample.

We detected the temperature-sensitive {\deltabfz[OIII] 436.3 nm} line in the average spectra stacked in bins of stellar mass and redshift, allowing us to calculate their metallicities using the direct method.  The WiggleZ galaxies have similar metallicities to normal emission-line galaxies at intermediate stellar masses ($8.8<\log(M_*/M_\odot)<10$), but they have significantly lower metallicities at high stellar masses ($10<\log(M_*/M_\odot)<12$). This is not an effect of evolution as the WiggleZ metallicities do not vary with redshift; it is most likely a property specific to the extremely UV-luminous WiggleZ galaxies, although similar behaviour has been reported for other galaxy types in the literature \citep[e.g.][]{Hunt2012}.


\section*{Acknowledgements}

This project would not be possible without the superb AAOmega/2dF facility provided by the Anglo-Australian Observatory. We wish to
thank all the AAO staff for their support, especially the night assistants, support astronomers and Russell Cannon (who greatly
assisted with the quality control of the 2dF system).

We also wish to thank Heinz Andernach for feedback on this manuscript; Alejandro Dubrovsky for writing software used to check the guide star and blank sky positions; Maksym Bernyk, David Barnes and Rod Harris for help with the database construction; Peter Jensen for assistance with the redshift measurements; Michael Stanley for help with the selection of new {\it GALEX} positions. We thank the referee for many suggestions which have improved the manuscript.

%

We wish to acknowledge financial support from The Australian Research
Council (grants DP0772084, DP1093738 and LX0881951 directly for the
WiggleZ project, and grant LE0668442 for programming support),
Swinburne University of Technology, The University of Queensland, the
Anglo-Australian Observatory, and The Gregg Thompson Dark Energy
Travel Fund. MJD thanks the International Centre for Radio Astronomy
Research, University of Western Australia
for travel support. SB acknowledges funding support from the Australian 
Research Council through a Future Fellowship (FT140101166).

{\it GALEX} is a NASA 
Small Explorer, launched in April 2003. We gratefully acknowledge NASA's support for construction, operation and 
science analysis for the {\it GALEX} mission, developed in cooperation with the Centre National d'Etudes Spatiales of 
France and the Korean Ministry of Science and Technology. 

Funding for the SDSS and SDSS-II has been provided by the Alfred
P. Sloan Foundation, the Participating Institutions, the National
Science Foundation, the U.S. Department of Energy, the National
Aeronautics and Space Administration, the Japanese Monbukagakusho, the
Max Planck Society, and the Higher Education Funding Council for
England. The SDSS Web Site is \url{http://www.sdss.org/}.

The RCS2 survey is based on observations obtained with 
MegaPrime/MegaCam, a joint project of the Canada-France-Hawaii Telescope (CFHT) and CEA/DAPNIA, at the
CFHT which is operated by
the National Research Council (NRC) of Canada, the Institut National
des Sciences de l'Univers (CNRS) of France, and the University
of Hawaii.  The RCS2 survey is supported by grants to HKCY from
the Canada Research Chair program and the Discovery program of
the Natural Science and Engineering Research Council of Canada.






\bsp	
\label{lastpage}
\end{document}